\documentclass[12pt,a4paper]{article}

\usepackage{epsfig,graphicx,bbm}
\usepackage{amsmath,amsfonts,amssymb}
\usepackage{citesort}
\usepackage[footnotesize]{caption}
\usepackage{psfrag}
 
\newcommand{\unit}{\leavevmode\hbox{\small1\kern-3.6pt\normalsize1}}

\allowdisplaybreaks

\psfrag{mN3}{${ \bf m_{N_3}}$}
\psfrag{modteta2}{${\bf |\theta_2|}$} 

\begin{document}

\begin{flushright}
FTUAM-08/24\\
IFT-UAM/CSIC-08-84\\
IFIC/09-12\\

\vspace*{5mm}
\end{flushright}

\vspace*{2mm}
\begin{center}
{\Large \textbf{Sensitivity to the Higgs Sector \\[1.5mm] 
of SUSY-Seesaw Models in the \\[3mm]
    Lepton Flavour Violating $\tau \to \mu f_0(980)$ decay} }

\vspace{1.0cm} 
{\large
M.~J.~Herrero\,$^{a}$, J.~Portol\'es\,$^{b}$ 
and A.~M.~Rodr\'{\i }guez-S\'anchez\,$^{a}$}\\[0.4cm]

{$^{a}$\textit{Departamento de F\'{\i }sica Te\'{o}rica 
and Instituto de F\'{\i }sica Te\'{o}rica, IFT-UAM/CSIC \\[0.1cm]
Universidad Aut\'{o}noma de Madrid,
Cantoblanco, E-28049 Madrid, Spain}}\\[0.3cm]

{$^{b}$\textit {IFIC, Universitat de Val\`encia - CSIC, Apt. Correus 22085, E-46071 Val\`encia, Spain}}\\[0.1cm]

\vspace*{2cm} 

\begin{abstract}

\vspace*{0.5cm} 
In this work we study the Lepton Flavour Violating semileptonic 
$\tau \to \mu f_0(980)$ decay within the context of SUSY-Seesaw Models, where the
MSSM spectrum is extended by three right handed neutrinos and their SUSY partners,  
and where the seesaw mechanism is used to generate the neutrino masses. 
 We estimate its decay rate when it proceeds via the 
 Higgs mediated channel $\tau \to \mu H^* \to \mu f_0(980)$, where $H$ refers
 to the CP-even 
MSSM Higgs bosons $h^0$ and
$H^0$, and the 
Lepton Flavour Violating $\tau \mu H$ vertex is radiatively generated via 
SUSY loops. In order to describe the $f_0(980)$ meson we follow the guidelines
from chiral constraints. As an implication of our computation,   
we explore the
sensitivity to the Higgs sector in this decay and compare it 
with other LFV tau decay channels. The confrontation of our predictions 
for BR$(\tau \to \mu f_0(980))$ with its very competitive 
present experimental bound leads us to extract some interesting restrictions on the 
most relevant model parameters, particularly, $\tan \beta$ and $m_{H^0}$.             
\end{abstract}

\end{center}

\newpage 
\section{Introduction}\label{intro}
The study of Lepton Flavour Violating (LFV) processes provides 
one of the most efficient indirect tests of supersymmetric (SUSY) extensions of the 
Standard Model of Particle 
Physics~\cite{Borzumati:1986qx,Hisano:1995nq,Hisano:1995cp,Hisano:1998fj,Kuno:1999jp,Casas:2001sr}.
  The reason is because
	in SUSY models the lepton and slepton mass matrices are not diagonal in
	flavour simultaneously, and  this misalignment 
	leads to intergenerational interactions between leptons and sleptons 
	with neutralinos and charginos at tree level, that when placed into 
	the loops of lepton 
	flavour changing processes, can
	generate large rates. Furthermore, in the case of SUSY-Seesaw
	models, with extended lepton and slepton sectors by three right
	handed neutrinos, $\nu_R$, and their SUSY partners, $\tilde \nu_R$, 
	and where the seesaw
	mechanism is used to generate the neutrino masses (i.e., the 
	so called Seesaw models of type I~\cite{seesaw:I}), the size of the 
	off-diagonal (in flavour) slepton mass matrix elements that are 
	responsible for LFV, 
	is governed by the
strength of the neutrino Yukawa couplings which can be $Y_\nu \sim {\cal O}(1)$ 
or even larger for heavy $M_{\nu_R} \sim 10^{14}-10^{15}$ GeV.
Thus, an interesting connection between neutrino and 
LFV physics follows, because the large Yukawa couplings of the 
Majorana neutrinos induce, 
via loops of SUSY particles, important contributions to LFV processes. 
In fact, these contributions are in some
cases~\cite{Borzumati:1986qx,Hisano:1995nq,Hisano:1995cp,Hisano:1998fj,Kuno:1999jp,Casas:2001sr,Babu:2002et,Sher:2002ew,Dedes:2002rh,Kitano:2003wn,Brignole:2003iv,Brignole:2004ah,Arganda:2004bz,Paradisi:2005fk,Arganda:2005ji,Paradisi:2005tk,Chen:2006hp,Antusch:2006vw,Arganda:2007jw,Arganda:2008jj,Fukuyama:2005bh,Arganda:2008yc}, at the reach 
of the present experimental sensitivity~\cite{Amsler:2008zzb}.

The LFV process that is the most sensitive to the neutrino Yukawa couplings, in the 
SUSY-Seesaw 
context, is $\mu \to e \gamma$, where the present experimental sensitivity 
is at $1.2\times 10^{-11}$~\cite{Brooks:1999pu,Ritt:2006cg}. Also $\mu-e$
conversion in heavy nuclei, with present bounds at 
CR($\mu-e,{\rm Ti})<4.3\times 10^{-12}$~\cite{Dohmen:1993mp} and 
CR($\mu-e,{\rm Au})<7\times 10^{-13}$)~\cite{Bertl:2006up},  
and $\mu \to 3 e$ with BR$(\mu \to 3 e) < 1.0 \times
10^{-12}$~\cite{Bellgardt:1987du}, 
are quite sensitive to LFV in the $\mu-e$ 
sector. 
 The most competitive LFV process in the $\tau-\mu$ sector is 
 $\tau \to \mu \gamma$, whose upper bound  is now set to 
 $1.6\times
 10^{-8}$~\cite{Aubert:2005ye,Hayasaka:2007vc,Banerjee:2007rj}.
Moreover, the sensitivity to LFV in $\tau \rightarrow 3\mu$ has improved
remarkably in the last years. The present upper bounds from BELLE and BABAR 
collaborations are $3.2\times 10^{-8}$ and $5.3\times 10^{-8}$, 
respectively~\cite{Miyazaki:2007zw,Aubert:2007pw}. In the last years, 
several interesting bounds at the $10^{-8}$ level for some LFV semileptonic
tau decays have also
been provided~\cite{Yusa:2006qq,Aubert:2006cz,Miyazaki:2007jp}. 

  In this work, we study the LFV semileptonic tau decay channel 
  $\tau \to \mu f_{0}(980)$ , which is competitive with other LFV 
  tau decays due to the  recently reported bound 
  by BELLE collaboration~\cite{Miyazaki:2008mw},
 $BR(\tau \to \mu f_{0}(980))\times
BR(f_{0}(980)\to \pi^{+}\pi^{-})< 3.4\times10^{-8}$. 
In fact, it is at present, the best bound 
 in semileptonic LFV tau decays, improving the
other present competitive bound of 
$BR(\tau \to \mu \eta)< 5 \times 10^{-8}$~\cite{Banerjee:2007rj}.
   The advantage of $\tau \to \mu \eta$~\cite{Sher:2002ew,Brignole:2004ah,Arganda:2008jj}  and 
   $\tau \to \mu f_{0}(980)$~\cite{Chen:2006hp} over the 
   $\tau\to \mu \gamma$ channel is their potential sensitivity to the Higgs sector. 
   Whereas the $\tau \to \mu \eta$ can be mediated by a $Z$ boson and a CP-odd 
   Higgs boson $A^0$, and it is dominated by the $A^0$ just at large
   $\tan \beta \gtrsim 20 $~\cite{Arganda:2008jj,Arganda:2008yc}, 
   the $\tau \to \mu f_{0}(980)$ decay is exclusively mediated by the exchange of 
   the neutral CP-even Higgs bosons $H^{0}$ and $h^{0}$. Therefore, through 
   the $\tau \to \mu f_{0}(980)$ channel one is testing directly 
   the neutral CP-even Higgs sector at all $\tan \beta$ values .

 Our computation of the BR$(\tau \to \mu f_{0}(980))$ improves the
 estimate of~\cite{Chen:2006hp} in several aspects. First, we demand
 compatibility with present data on light neutrino masses and mixings. 
  Second, we do not use the mass insertion approximation, we take into account
 the full set of  SUSY one-loop diagrams in the LFV vertex $\tau \mu H$ 
 ($H=h^0,H^0$), and include the two contributions mediated by the 
$h_{0}$ and $H_{0}$ respectively. Consequently, we explore the full 
$5 \leq \tan\beta\leq  60$ interval. Besides,
the hadronization of quark 
bilinears into the
$f_0$(980) meson is performed here quite differently than 
in \cite{Chen:2006hp}, where a simplified quark-flavour scheme was used
to express these
bilinears in terms of phenomenological meson decay constants.
We instead pay close attention to the chiral constraints, following the standard Chiral 
Perturbation Theory
($\chi$PT)~\cite{Weinberg:1978kz,Gasser:1983yg,Gasser:1984gg} and the
Resonance Chiral Theory
(R$\chi$T)~\cite{Ecker:1989yg,Ecker:1988te,Pich:2002xy,Cirigliano:2003yq,Cirigliano:2006hb} to incorporate 
resonances. Concretely, we follow the description of $f_0(980)$ in \cite{Cirigliano:2003yq}, 
where it is defined by a mixing between the octet and singlet components 
of the nonet of the 
scalar resonances which are included in R$\chi$T. 
Furthermore, we do not work in a generic Minimal Supersymmetric
Standard Model (MSSM) 
but in constrained models with input parameters set at the high energies.
Concretely we focus on    
 two particular constrained SUSY
 scenarios of remarkable interest: the usual constrained MSSM 
 (CMSSM) scenario~\cite{Kane:1993td}, with universal soft SUSY masses 
 at the gauge coupling unification scale, and the so-called Non-Universal 
 Higgs Mass (NUHM) scenario~\cite{Ellis:2002iu}, with all the scalar 
 soft masses being universal except for the Higgs sector ones. In 
 this later case the physical Higgs boson masses, $m_{h^0}$ and $m_{H^0}$, can be 
 both light, $\sim 100-250$ GeV, indeed 
 close to their present experimental lower bounds and, therefore, the corresponding 
 Higgs mediated 
 contribution to the previous LFV processes can be relevant, even for large 
 soft SUSY masses at $\sim {\cal O}(1\,{\rm TeV})$. This is precisely the main interest of the channel 
 $\tau \to\mu f_{0}(980)$, namely, the fact that the decay 
rates can be sizeable even for large SUSY masses, 
$M_{\rm SUSY} \sim {\cal O}(1\,{\rm TeV})$, 
in clear contrast with 
other competitive tau flavour violating channels like 
$\tau \to \mu \gamma$, whose rates decrease as $1/M^2_{\rm SUSY}$
 and lay below the present experimental bound for such a heavy SUSY spectrum.

 \section{Framework for the $\tau \to \mu f_0(980)$ decay}\label{framework}
For the present study of  the $\tau \to \mu f_0(980)$ decay, we choose a 
SUSY-Seesaw framework  where the spectrum of the MSSM is 
 enlarged by three right-handed neutrinos, $\nu_{R_i}$ ($i=1,2,3$), 
 and their SUSY 
 partners, $\tilde \nu_{R_i}$ ($i=1,2,3$). Here 
we  assume a seesaw mechanism
for neutrino mass generation and use, in particular, 
the parameterization proposed in~\cite{Casas:2001sr} where
the solution to the seesaw equation, relating the parameters of 
the six physical (mass eigenstates) Majorana neutrinos, $\nu_i$ , and $N_i$ $(i=1,2,3)$ to the 
neutrino Yukawa
couplings, is written as 
\begin{equation}\label{seesaw:casas}
m_D\,=Y_\nu\, v_2=\, i \sqrt{m^\text{diag}_N}\, R \,
\sqrt{m^\text{diag}_\nu}\,  U^\dagger_{\text{PMNS}}\,.
\end{equation}
Here, the Dirac mass, $m_D$, the Yukawa neutrino coupling, $Y_\nu$, and $R$ are 
$3 \times 3$ matrices with full structure in flavour space. The orthogonal 
matrix $R$ is defined by three complex angles $\theta_i$
($i=1,2,3$)~\cite{Casas:2001sr}.  
$m_{\nu}^\mathrm{diag}=\, \mathrm{diag}\,(m_{\nu_1},m_{\nu_2},m_{\nu_3})$ denotes the
three light neutrino masses, and  
$m_N^\mathrm{diag}\,=\, \mathrm{diag}\,(m_{N_1},m_{N_2},m_{N_3})$ the three heavy
ones. The two Higgs vacuum expectation values are
 $v_{1(2)}= \,v\,\cos (\sin) \beta$, with $v = 174$ GeV. The
 Pontecorvo-Maki-Nakagawa-Sakata unitary matrix 
 $U_{\rm PMNS}$~\cite{Pontecorvo:1957cp,Maki:1962mu} is given by
the three (light) neutrino mixing angles $\theta_{12},\theta_{23}$ and $\theta_{13}$, 
and three phases, $\delta, \phi_1$ and $\phi_2$. With this 
parameterization is easy to accommodate
the neutrino data. It further allows for large
Yukawa couplings $Y_\nu \sim \mathcal{O}(1)$ by
choosing large entries in $m^{\rm diag}_N$ and/or $\theta_i$. 

For the numerical predictions in this work we will 
set:
\begin{align}
& m_{\nu_1}^2 \,\simeq\, 0 \,,\quad  \; \; \; m_{\nu_2}^2\,=\,
\Delta\, m^2_\text{sol} \,=\,8\,\times 10^{-5}\,\,\text{eV}^2\,, \; \; \; 
\quad  m_{\nu_3}^2\,=\,
\Delta \, m^2_\text{atm} \,=\,2.5\,\times 10^{-3}\,\,\text{eV}^2\,,
\nonumber \\[3mm]
& 
\theta_{12}\,=\,30^\circ\,, 
\quad \;\; \; 
\theta_{23}\,=\,45^\circ\,,
\quad \; \; \; 
\theta_{13}\,=\,5^\circ\,,
\quad \quad \; \; \; 
\delta\,=\,\phi_1\,=\,\phi_2\,=\,0\,,
\label{light}
\end{align}
which are compatible with present neutrino data~\cite{Amsler:2008zzb},
and consider the two possibilities for the heavy neutrinos: 1) Degenerate,
with $m_{N_1}=m_{N_2}=m_{N_3}\equiv m_{N}$; and 2) Hierarchical, with
 $m_{N_1}<<m_{N_2}<<m_{N_3}$. This later case is well known to provide a
 plausible scenario for the Baryon Asymmetry of the Universe (BAU) 
 via leptogenesis.

Regarding the SUSY parameters we will work within two different constrained
MSSM-Seesaw scenarios, 
the CMSSM
with universal soft SUSY breaking parameters (including the extended sneutrino
sector) and the NUHM model with
 non-universal Higgs soft masses. Thus, in addition to the previous neutrino
 parameters, $m_{N_i}$ and $\theta_i$, the input parameters of these two models 
 are respectively,
\begin{align}
\text{CMSSM :} \ & M_0\,, M_{1/2}\,,A_0\,, \tan \beta\,, 
\text{sign}(\mu)\,,
\nonumber \\
\text{NUHM :}\ & M_0\,, M_{1/2}\,,A_0 \,, \tan \beta\, ,
\text{sign}(\mu)\,,M_{H_1}^2\,=\,M^2_0\,(1+\delta_1),
M_{H_2}^2\,=\,M^2_0\,(1+\delta_2).
\end{align}
where $M_0$, $M_{1/2}$ and $A_0$ are the universal soft SUSY breaking scalar masses, gaugino
masses and trilinear couplings, respectively, at the gauge
coupling unification scale, $M_X \simeq 2 \times 10^{16}$ GeV. The other
parameters are, as usual, the ratio of the two Higgs vacuum expectation values,
$\tan \beta=v_2/v_1$, and the sign of the $\mu$ parameter, $\text{sign}(\mu)$.
Notice, that the departure from universality in the soft Higgs masses of the NUHM 
is parameterized here 
in terms of the two dimensionless parameters $\delta_1$ and $\delta_2$. Consequently, 
by taking $\delta_1 =\delta_2=0$ in the NUHM one recovers the CMSSM case. 
Finally,  
 in order to evaluate the previous SUSY parameters and the physical masses 
 at low energies (taken here as 
the Z gauge boson mass $m_Z$), we solve the full one-loop Renormalization Group 
Equations
(RGEs) including the extended neutrino and sneutrino sectors. For this and 
the computation of the full spectra at the low energy we use here
the public FORTRAN code SPheno~\cite{Porod:2003um}. In the numerical estimates
we will
set $M_0= M_{1/2}$, $A_0=0$ and $\text{sign}(\mu)=+1$, for simplicity.

\begin{figure}[ht!]
   \begin{center} 
     \begin{tabular}{cc} \hspace*{-12mm}
  	 \psfig{file=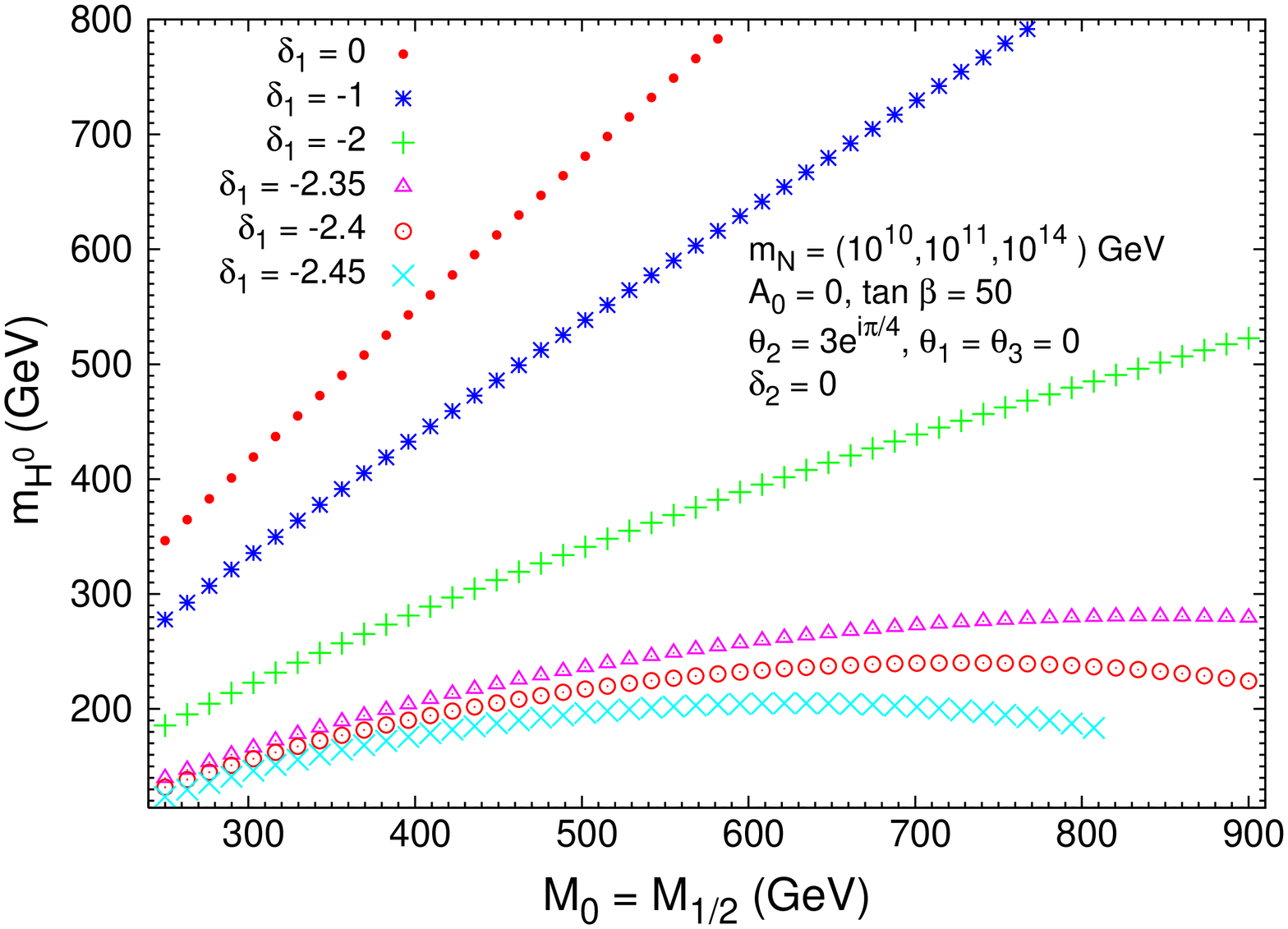,width=60mm,angle=270,clip=}
  &
  	 \psfig{file=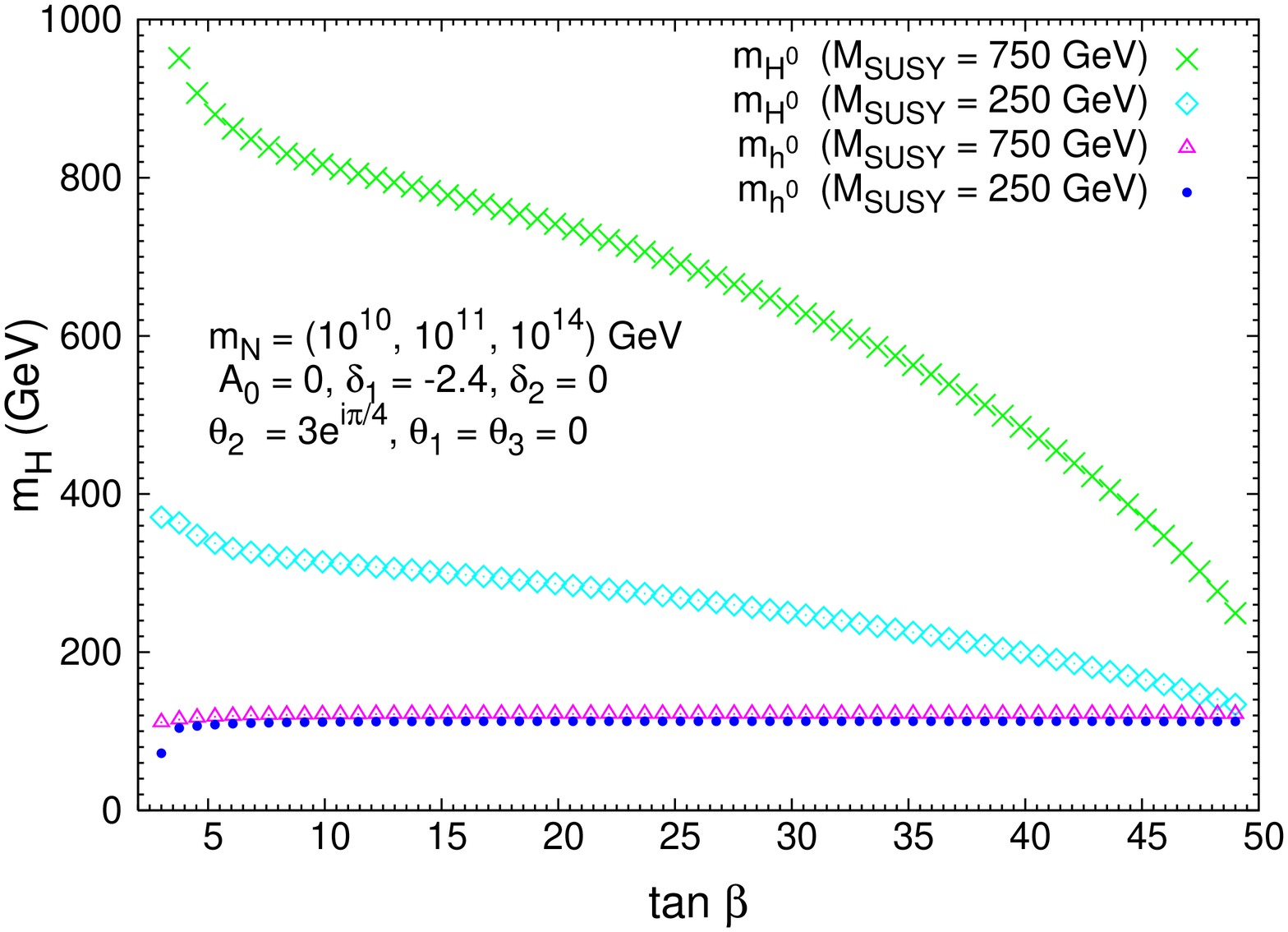,width=60mm,angle=270,clip=}
     \end{tabular}
     \caption{CP-even Higgs boson masses in the NUHM-Seesaw scenario: 1) $m_{H^0}$  as a function of
     $M_{\text{SUSY}} = M_0 = M_{1/2}$ for several input $\delta_{1,2}$ (left
     panel). The predictions in the CMSSM-Seesaw 
     scenario ($\delta_1 = \delta_2 = 0$) are included for comparison; 2) $m_{H^0}$ and  $m_{h^0}$ as 
     functions of $\tan \beta$ for $M_{\text{SUSY}}=$ 250 GeV and
     750 GeV (right panel).}\label{fig:mH_CMSSM-NUHM} 
   \end{center}
 \end{figure}

For the purpose of the present analysis the most relevant difference
between the two previous constrained SUSY-Seesaw scenarios is the
spectrum of the Higgs sector. In particular, we want to explore the
interesting case where the neutral Higgs bosons that mediate the
$\tau \to \mu f_0(980)$ decay are light, while keeping the SUSY spectra
heavy enough as to suppress the other competitive LFV tau decay 
channels like, for instance, $\tau \to \mu \gamma$. This is clearly possible within the
NUHM-Seesaw scenario, as illustrated in Fig.~\ref{fig:mH_CMSSM-NUHM}.   
 We see in this figure that, by properly adjusting the input $\delta_1$
and $\delta_2$ parameters, the heavy Higgs boson $H^0$ can get masses as low as 100-250
GeV even for a very heavy SUSY spectrum. For instance, for $\delta_1=-2.4$, $\delta_2=
0$, $\tan \beta= 50$, $M_{\rm SUSY}=M_0=M_{1/2}=750$ GeV and the other input 
parameter values as specified 
in this figure, we get $m_{H^0}=249$ GeV and $m_{h^0}=122$ GeV,
 to be compared with $m_{H^0}=998$ GeV and $m_{h^0}=122$ GeV  of the
CMSSM-Seesaw case. With other specific choices for $\delta_2 \neq  0$ one gets even lower values
of $m_{H^0}$~\cite{Arganda:2008jj}. For the following numerical analysis and, for simplicity, we will set,
however, 
$\delta_2=0$ and play just with $\delta_1$. It is worth also mentioning that
 the predictions for $m_{A^0}$ (not shown in this figure) 
are practically indistinguishable from those of $m_{H^0}$~\cite{Arganda:2008jj}.          

\begin{figure}[ht!]
   \begin{center} 
     \begin{tabular}{cc} \hspace*{-12mm}
  	\psfig{file=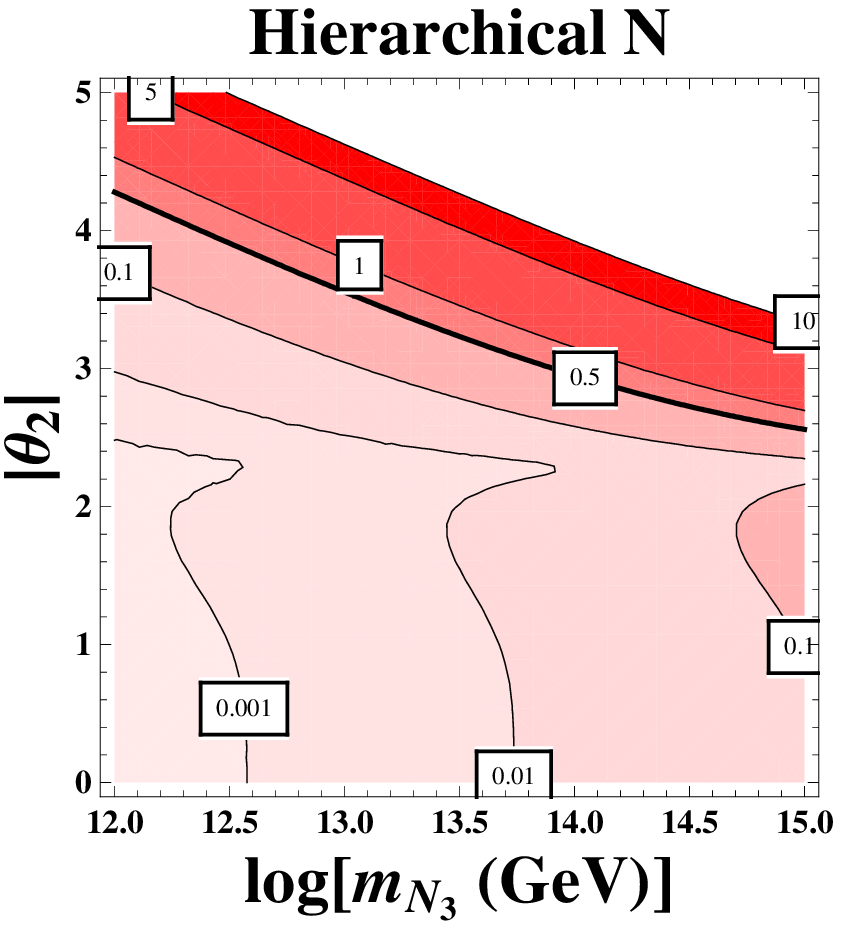,width=78mm,clip=}
  &
  	\psfig{file=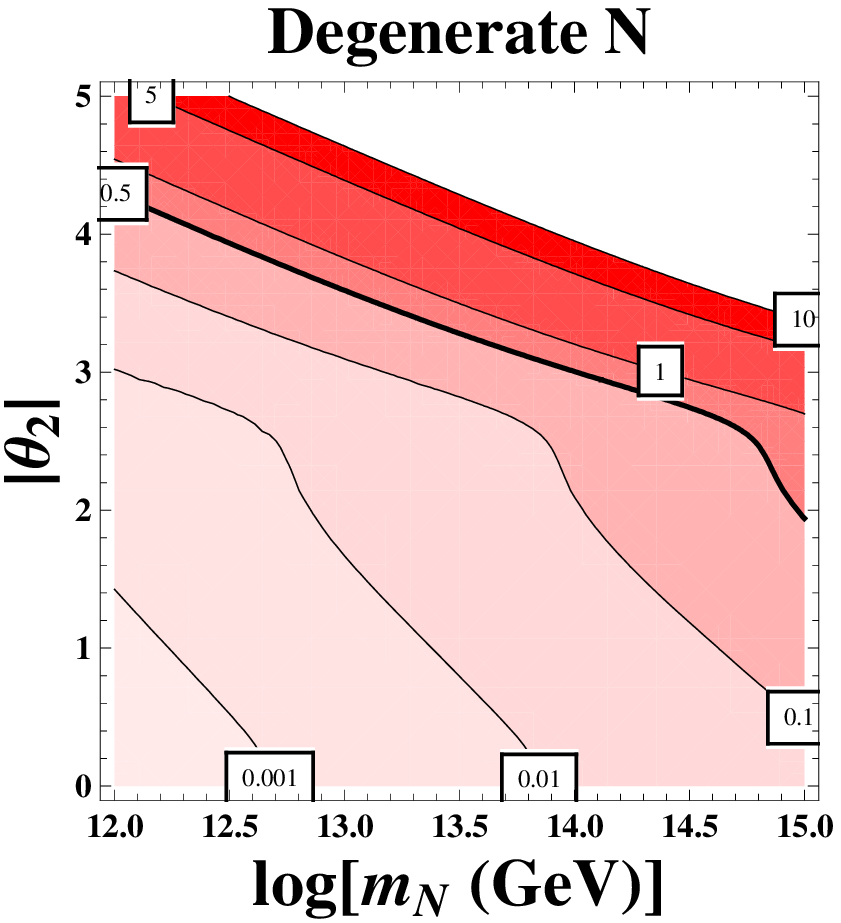,width=78mm,clip=} \\
\hspace*{-12mm}
\psfig{file=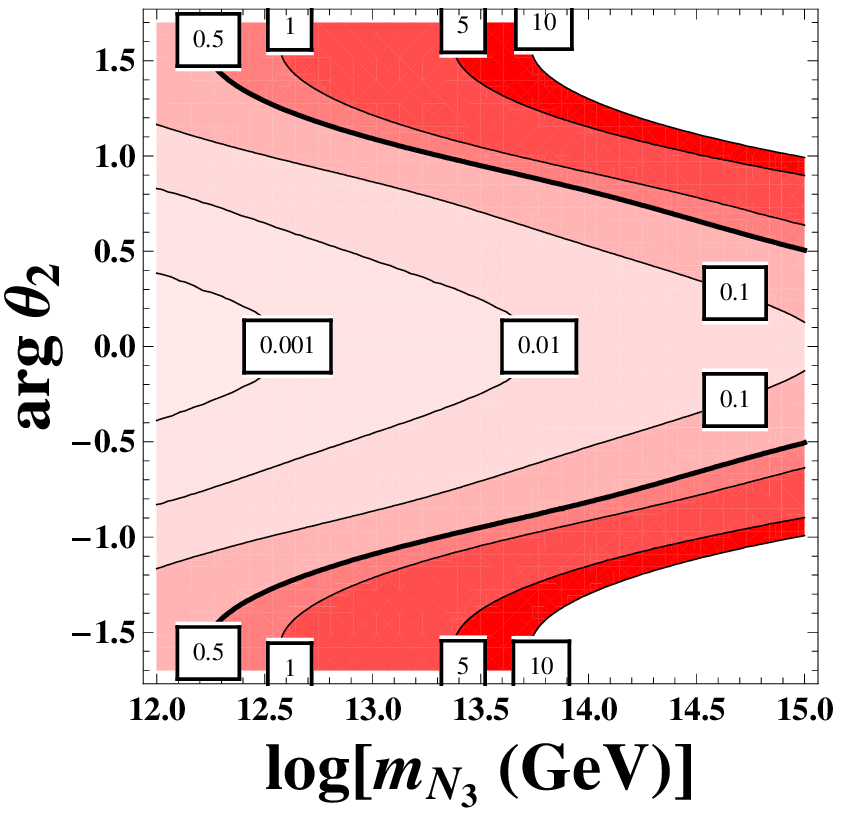,width=82mm,clip=}
&
\psfig{file=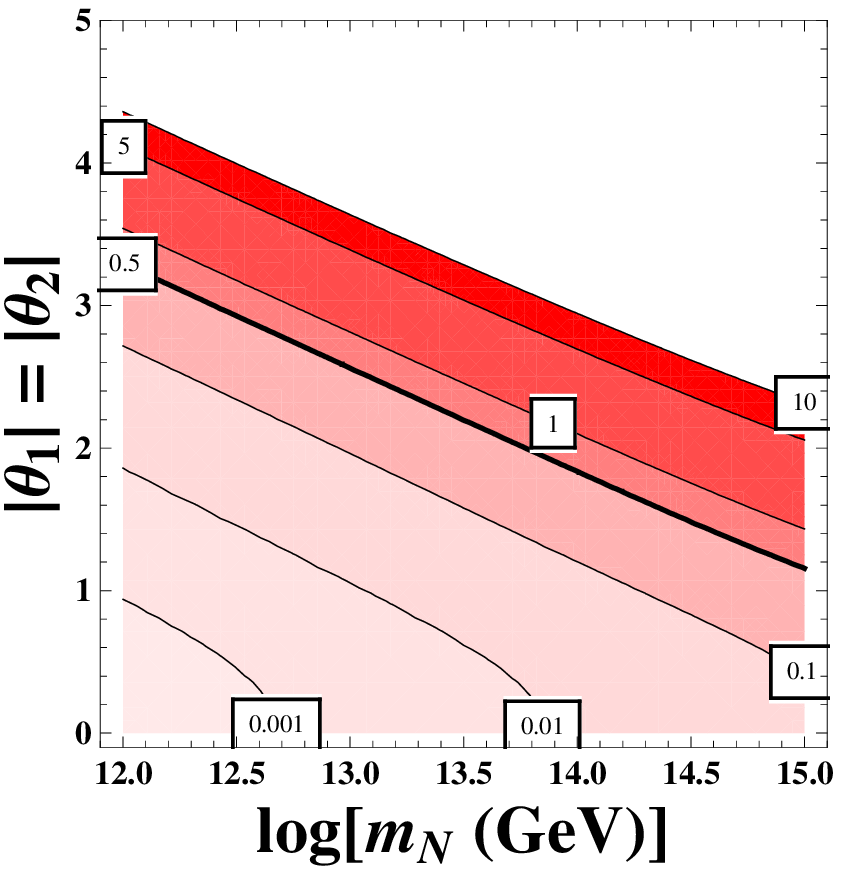,width=78mm,clip=}
     \end{tabular}
\caption{Contours of $|\delta_{32}|$ in the CMSSM-Seesaw scenario: 
1) For hierarchical heavy neutrinos. Upper left panel: 
in the ($|\theta_2|$,$m_{N_3}$) plane for ${\rm arg} \,\theta_2= \pi/4$. 
Lower left panel: in the 
     (${\rm arg}\, \theta_2$,$m_{N_3}$) plane for  $|\theta_2|=3$. The other heavy neutrino parameters 
     are set to $\theta_1=\theta_3=0$,
     $m_{N_1}=10^{10}$ GeV, $m_{N_2}= 10^{11}$ GeV; 2) For degenerate heavy neutrinos. Upper right
     panel: in the ($|\theta_2|$,$m_{N}$) plane for ${\rm arg}\,\theta_2= \pi/4$ and $\theta_1=\theta_3=0$. 
     Lower right panel: in the 
     ($|\theta_1|=|\theta_2|$,$m_{N}$) plane for   
     ${\rm arg} \,\theta_1= {\rm arg} \,\theta_2= \pi/4$, and $\theta_3=0$. In all
     plots we have set: $M_{\rm SUSY}=M_0=M_{1/2}$, $A_0=0$, $\tan \beta= 50$, 
     and the $\theta_i$
     are expressed in radians.}
 \label{fig:contours} 
   \end{center}
 \end{figure}

Within the previous scenarios for the neutrino and SUSY sectors, 
it is well known that one can get large LFV decay rates if one chooses 
large entries in $m^{\rm diag}_N$ and/or complex $\theta_i$, basically
due to the large size of $Y_\nu$ in these models. This can
be understood more easily in the Leading Logarithmic (LLog)
approximation where, the tau-muon flavour violation, which is of our
interest here, is qualitatively
well described by the parameter,
\begin{equation}
\label{delta32_Llog}
\delta_{32}\,=\,
-\frac{1}{8\, \pi^2}\, \frac{(3\, M_0^2+ A_0^2)}{M_{\rm SUSY}^2}\, (Y_{\nu}^\dagger\, 
L\, Y_{\nu})_{32}\,\,,\,\, \; \; \; 
L_{kl}\, \equiv \,\log \left( \frac{M_X}{m_{N_k}}\right) \,
\delta_{kl}\,,\, k,l=1,2,3,
\end{equation}
where $M_{\rm SUSY}$ is an average SUSY mass.
The size of $|\delta_{32}|$ can be indeed quite large. For instance, 
for mass values of 
the heavy neutrinos $m_{N_3}$ (or $m_N$) in the range
$10^{14}-10^{15}$ GeV and $\theta_i$ (i= 1 or/and 2) with large modulus 
in the range $3-5$ or/and large argument in the range $\left[ \pm \pi/4 , \pm \pi/2\right]$
one can get values of $|\delta_{32}|$ as large as 0.5-10. 
This is clearly illustrated in the contour plots of
Fig.~\ref{fig:contours}, where we have considered both scenarios with
either degenerate or hierarchical heavy neutrinos and we have explored
in the $(m_{N_i},\theta_i)$ parameter space. In the hierarchical case the
relevant mass is the heaviest one $m_{N_3}$ and the predictions for 
$|\delta_{32}|$ do not vary appreciably with $m_{N_{1,2}}$. In addition,
we have checked that $|\delta_{32}|$ is nearly constant with $\theta_3$.
The contour
plots for $\theta_1$ (not shown) are very similar to those of 
$\theta_2$. We have also found that the largest values of 
$|\delta_{32}|$ are obtained for the degenerate case with both $\theta_1$
and $\theta_2$ being large. This is also clearly illustrated in the lower right panel
of Fig.~\ref{fig:contours}. For
instance, we get $|\delta_{32}| \simeq 5$ for $m_N=10^{14}$ GeV and 
$\theta_1=\theta_2=3\,{\rm exp}\,(i\pi/4)$.
Notice also that values of $|\delta_{32}|$ larger than  
$\sim 0.5$ correspond in our parameterization of the Yukawa
coupling matrices in (\ref{seesaw:casas}) 
to values of $|Y_\nu|^2/(4\pi)$  
that are above the threshold where the 
SPheno code sets the limit of perturbativity, which is at
$|Y_\nu|^2/(4\pi) \sim 1.5$. It means that, in the following, 
we will be able to provide full
predictions for the decay rates with the SPheno code only for 
those model parameters producing $Y_\nu$ 
values that are within the
perturbativity region or, equivalently, leading to 
$|\delta_{32}| < 0.5$. The implications for the 
$\tau \to \mu f_0(980)$ decay 
of values $|\delta_{32}| \geq 0.5$ will be explored later, 
not with our full computation implemented by us in SPheno, 
but using an approximate formula that will also be presented here
and that turns out to work reasonably well. 

Next, we specify our framework for the hadronization of the quark bilinears into
the $f_0(980)$ meson.  We use here the chiral 
Lagrangian of R$\chi$T that is a suitable tool to realise the $1/N_C$
expansion of $SU(N_C)$ QCD and includes both the Goldstone bosons $\Phi$($\pi$,
$K$ and $\eta$) and the resonances as active degrees of freedom, 
and their interactions. For the present work, it
is sufficient to consider the lightest nonet of scalar resonances $R(0^{+})$  
in R$\chi$T,
\begin{eqnarray}\label{eq:rcht}
{\cal L}_{R\chi T} & = & {\cal L}_{\chi}^{(2)} \, + \, {\cal L}_{\rm kin}^R \, +
\, {\cal L}_{(2)}^R \; , 
\end{eqnarray}
where,
\begin{eqnarray}\label{eq:lagro}
 {\cal L}^{(2)}_{\chi} & = & \frac{F^2}{4} \, \langle u_{\mu} \, u^{\mu} \, + 
 \, \chi_{+} \rangle \,\,,\,\, \; \; \; F \simeq F_\pi \simeq 92.4\,{\rm MeV},\nonumber \\
{\cal L}^{R}_{\rm kin}  & = &   \frac{1}{2}\langle \nabla^{\mu}R \nabla_{\mu}R 
- M_{R}^{2} R^2 \rangle \, 
 \, , 
\nonumber \\
{\cal L}_{(2)}^R & = &  \, c_{d} \, \langle \, 
R \, u_{\mu} u^{\mu} \, \rangle \, + \, c_{m} \, \langle \, R \, \chi_+ \,
\rangle \; ,
\end{eqnarray}
and $\langle ... \rangle$ is short for a trace in the flavour space. The other
quantities in 
(\ref{eq:lagro}) are:
\begin{eqnarray}
 u_{\mu} & = & i [ u^{\dagger}(\partial_{\mu}-i r_{\mu})u-
u(\partial_{\mu}-i \ell_{\mu})u^{\dagger} ] \ , \; \; \; \;  \; \; \; \; \; \; \; \; \; \; \; \; 
u = \exp   \, [ i\,\Phi /(\sqrt{2} F)\,] \, , 
\nonumber \\ 
\chi_{+} & = & u^{\dagger}\chi u^{\dagger} + u\chi^{\dagger} u\  
, \; \; \; \; \; \; \; \; \; \; \; \; \; \; \; \; \; \; \; \; \; \;\; \; \; \; \; \; \; 
\; \; \; \; \; \; \; \; \; \; \; \; \; \; 
\chi=2B_0(s+ip) \;  ,  \nonumber \\ 
\nabla_\mu R &	= &	 \partial_{\mu} R	 +
[\Gamma_{\mu}, R] \;, \; \; \; \;  \; \; \; \; \; \; \; \; \; \; \; \; \; \; \; \; \; 
\Gamma_\mu  =  \frac{1}{2} \, [ \,
u^\dagger (\partial_\mu - i r_{\mu}) u +
u (\partial_\mu - i \ell_{\mu}) u^\dagger \,], 
\end{eqnarray}
being $\Phi$ the nonet of Goldstone bosons, $\ell_\mu = v_\mu - a_\mu$, 
$r_\mu = v_\mu + a_\mu$, and $v_\mu$, $a_\mu$, $s$ and $p$ are the nonets of 
vector, axial-vector, scalar and pseudoscalar external fields, respectively.
Short-distance dynamics \cite{Pich:2002xy} constraints the couplings of R$\chi$T by imposing
the QCD ruled behaviour of Green functions and associated form factors. For the couplings
in ${\cal L}_{(2)}^R$ one gets\footnote{Short-distance constraints on the R$\chi$T couplings
depend on the operators included. The result in (\ref{constantes}) is obtained when only
linear operators in the resonances are considered \cite{Ecker:1988te}. A weaker constraint, though
compatible with that result, arises if non-linear couplings in the resonances are included
\cite{Cirigliano:2006hb}.}~:
\begin{eqnarray}
2 \, c_m & = & 2 \, c_d \; = \, F \;.
\label{constantes}
\end{eqnarray}
Finally the chiral tensor $\chi$ gives masses to the Goldstone bosons through the external scalar field.
In the isospin limit one has~:
 \begin{eqnarray}
2 \,  B_0 \, m_u & = & 2 \, B_0 \, m_d \; = \, m_{\pi}^2 \; , \nonumber \\
2 \, B_0 \, m_s & = & 2 \, m_K^2 - \, m_{\pi}^2  \, .
\label{mq-mP}
\end{eqnarray} 
The QCD spectrum of scalar resonance states is far from being settled and constitutes, at present,
a highly debated issue. It is not our goal in this article to enter in the details of the 
discussion and, therefore, we will attach to the scheme put forward in ~\cite{Cirigliano:2003yq}
for the description of the isosinglet $f_0(980)$ state. 
The later is defined as a rotation of the octet $R_8$ and the
singlet $R_0$ components of the $R(0^{+})$ nonet,
\begin{equation}
 \left( \begin{array}{c}
        R_8 \\
        R_0
        \end{array}
\right)
= 
\left( \begin{array}{cc}
        \cos \theta_S & \sin \theta_S \\
        - \sin \theta_S & \cos \theta_S 
       \end{array}
\right)
\left( \begin{array}{c}
        f_0(1500) \\
        f_0(980)
       \end{array}
\right) \, .
\label{rotation}
\end{equation}
The value of the $\theta_S$ mixing angle is uncertain. In the analysis carried out
in ~\cite{Cirigliano:2003yq}
considering nonet breaking
(i.e. subleading effects in the large-$N_C$ expansion) a
possible dual scenario is favoured~:
\begin{itemize}
\item[A)]  The candidates for the nonet are: $f_0(980)$,
$K^*_0(1430)$, $a_0(1450)$ and $f_0(1500)$. In this framework the $a_0(980)$ is dynamically generated
(through loops). The mixing angle, around
$\theta_S \simeq 30 ^\circ$, provides a dominant non-strange
component for the $f_0(980)$ state and, consequently, justifies
its dominant decay into two pions.
\item[B)] The nonet would be composed by: $f_0(980)$,
$a_0(980)$, $K^*_0(1430)$ and $f_0(1500)$. Hence $a_0(980)$ is a pre-existing state in the $N_C \to
\infty$ limit. The mixing angle in this
case is around $\theta_S \simeq 7^\circ$, that gives a noticeable strange 
component for the $f_0(980)$ state.
\end{itemize}
Given the uncertainty provided by the large corrections due to $1/N_C$
subleading effects we will consider the two previous scenarios for the 
$f_0(980)$ as plausible and will present estimates of the $\tau \to \mu f_0(980)$
decay rates for the two mixing angles, $\theta_S \simeq 7^\circ$
and $\theta_S \simeq 30^\circ$. The dispersion between these two results
can be considered as part of the theoretical error in our estimates. 

Finally, the hadronization of the relevant scalar quark bilinears 
into the $f_0(980)$ is implemented by replacing the following expressions in the results for the
decay rates at the quark level,
 \begin{eqnarray}
     \, \overline{u}  \, u & = & 
 - \left[ \frac{1}{2} S_3 + \frac{1}{2 \sqrt{3}} S_8 + \frac{1}{\sqrt{6}} S_0 \right] \, , \nonumber \\
   \, \overline{d} \, d & = & 
 - \left[ - \frac{1}{2} S_3 + \frac{1}{2 \sqrt{3}} S_8 + \frac{1}{\sqrt{6}} S_0 \right] \, , \nonumber \\
    \, \overline{s}  \, s & = & 
 - \left[ -\frac{1}{\sqrt{3}} S_8 + \frac{1}{\sqrt{6}} S_0 \right] \, , 
\label{quarkbilinears}
\end{eqnarray}
with  
\begin{eqnarray}
S_i & = & \frac{8}{\sqrt{2}} \, B_0 \, c_m \, R_i \, , \; \; \; \;  i=0,3,8\,,
\label{scalarcurrents}
\end{eqnarray}
and, according to (\ref{constantes}), $c_m =F/2$. 
As $R_3$ does not contain information on $f_0(980)$ (in the isospin limit) we will discard the 
$S_3$ contribution. 
\par
Before proceeding a word of caution is necessary when dealing with processes with resonances as
initial or final states. A resonance is not an asymptotic state as it decays strongly. Hence from
a quantum field theory point of view R$\chi$T only describes the creation, propagation and 
destruction of resonances and the later should not appear as \lq \lq in`` or \lq \lq out''
states. For instance, in our case the physical process should be $\tau \rightarrow \mu \pi \pi$
mediated by a $f_0(980)$ state, and not $\tau \rightarrow \mu f_0(980)$. Then it would proceed
to study the scalar state as was done with the vector ones in \cite{Arganda:2008jj}. However 
the description of scalars, as has been pointed out, is far from clear and therefore considering the
$f_0(980)$ as an asymptotic state should not increase effectively the already rather large
uncertainty.

\section{Results for BR$(\tau \to \mu f_0(980))$}

{\bf Analytical results}

\begin{figure}[hbtp]
  \begin{center} 
        \psfig{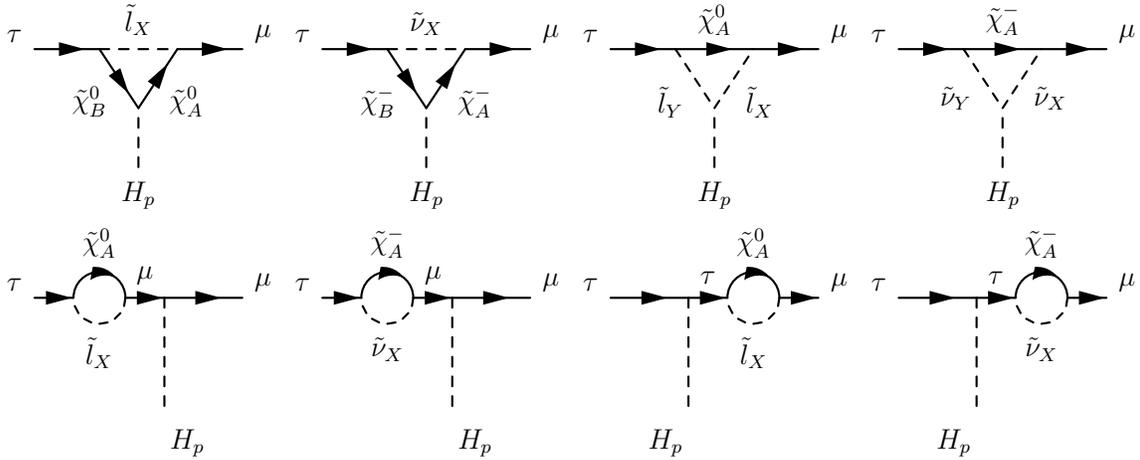}
    \caption{Relevant SUSY one-loop diagrams for the Higgs-mediated 
    contributions to the $\tau \to \mu f_0(980)$ decay. Here $H_p=h^0,H^0$.}\label{H_diagrams} 
  \end{center}
\end{figure}

The semileptonic  $\tau \to \mu f_0(980)$ decay  can be mediated by  $h^0$ and  
$H^0$ Higgs bosons, as shown in Fig.~\ref{LFVsemileptau_diagrams}. 
In this figure the LFV vertex is represented by a black circle and the 
hadronic vertex by a grey box.
The total amplitude for this 
  decay, $T_H=T_{h^0}+T_{H^0}$, is first evaluated at the quark level, that is
  for $\tau
\to \mu \overline{q} q$,  and then at the hadron level by substituting 
the quark bilinears by the corresponding scalar currents containing the 
$f_0(980)$ meson as evaluated from ${\cal L}_{R \chi T}$ in (\ref{eq:rcht}). 
The amplitude at the quark level can 
be computed in
terms of the corresponding $\tau \mu H_p$ one-loop 
vertex functions, $H_{L,R}^{(p)}$, with $H_p=h^0,H^0$, resulting from the
evaluation of the diagrams in Fig.~\ref{H_diagrams} with 
sleptons, ${\tilde l_X}$, sneutrinos, ${\tilde \nu_X}$, charginos, 
${\tilde \chi^-_A}$, and neutralinos, ${\tilde \chi^0_A}$,  in the loops.    
The resulting amplitude at the quark level is given by:
 \begin{figure}[h!]
   \begin{center} 
     \begin{tabular}{c} \hspace*{-12mm}
         \psfig{file=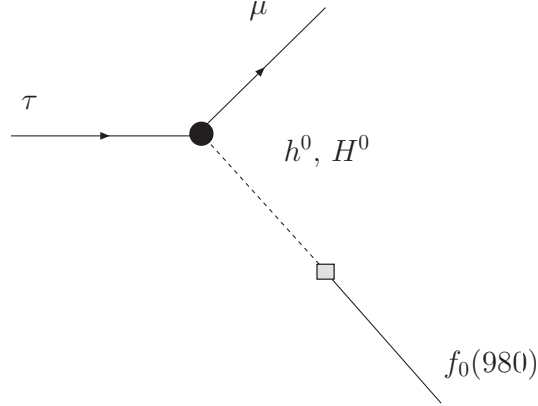,width=70mm,clip=} 
     \end{tabular}
     \caption{Higgs-mediated contributions to the LFV semileptonic $\tau \to \mu f_0(980)$ decay} 
     \label{LFVsemileptau_diagrams} 
   \end{center}
 \end{figure}  
\begin{eqnarray}
 T_H(\tau \to \mu \overline{q} q) & = & \sum_{h^0, H^0} \frac{1}{m_{H_p}^2} 
  \left\{ H_L^{(p)} S_{L,q}^{(p)}\left[ \overline{\mu} P_L 
\tau \right] \left[ \overline{q} P_L q \right]  \, + \, 
H_R^{(p)} S_{R,q}^{(p)} \left[ \overline{\mu} P_R 
\tau \right] \left[ \overline{q} P_R q \right] \right. \nonumber \\
& &  \; \; \; \; \; \; \; \; \left.  + \; H_L^{(p)} S_{R,q}^{(p)} 
\left[ \overline{\mu} P_L 
\tau \right] \left[ \overline{q} P_R q \right]  \, + \, 
H_R^{(p)} S_{L,q}^{(p)} \left[ \overline{\mu} P_R 
\tau \right] \left[ \overline{q} P_L q \right] \right\} \; .
\label{amplitude_quarks}
\end{eqnarray}
where $P_{L,R}=(1\mp\gamma_5)/2$, and  
 \begin{eqnarray}
 S_{L,q}^{(p)} &  = &  \frac{g}{2 m_W} \, \left(\frac{-\sigma_2^{(p)*}}{\sin
 \beta}\right) m_q \; , \; \; \; \;  q=u \, ; \nonumber \\
S_{L,q}^{(p)} & = & \frac{g}{2 m_W} \, \left(\frac{\sigma_1^{(p)*}}{\cos \beta}
\right) m_q\; , \;\; \, \; \;  q=d,s \, ; \nonumber \\
S_{R,q}^{(p)} & = &  S_{L,q}^{(p)*}
\label{Hqqcouplings}
\end{eqnarray}
with
\begin{eqnarray}
\sigma_1^{(p)} &=& \left( \begin{array}{c} \sin{\alpha} \\ -\cos{\alpha} \\ i
    \sin{\beta} \end{array} \right) \,,\,\,\,\, 
\sigma_2^{(p)}\, =\, \left( \begin{array}{c} \cos{\alpha} \\ \sin{\alpha} \\ -i
    \cos{\beta} \end{array} \right) \,. 
\end{eqnarray} 
Here $m_W$ is the $W$ gauge boson mass, $m_q$ is the $q$ quark mass, $\alpha$ 
is the mixing angle in the Higgs sector, and $g$ is the $SU(2)$ gauge coupling.
The three entries in $\sigma_{1,2}^{(p)}$ are, in order from top to bottom,
for $H_p = h^0, H^0, A^0$, respectively.

The results of the LFV vertex functions are taken from~\cite{Arganda:2004bz}, 
and are not written here explicitely for shortness. Just to mention that it
is a full one-loop computation, including all the contributions with charginos in the
loops, $H_{L(R),c}^{(p)}$, and those with neutralinos, $H_{L(R),n}^{(p)}$. 
Besides, 
all these contributions are written in terms of the physical particle masses. 
As we have 
  mentioned before,
  these physical masses are computed here in the SUSY-seesaw scenario by solving the
  one-loop RGEs with SPheno and for a given set of universal (in the CMSSM) or non-universal conditions 
  (for the NUHM) at 
  the unification scale. Since the three right-handed neutrinos and their SUSY
  partners are included in the RGEs, they will affect as well in the
  predicted physical masses at the low energies.   

To get the amplitude for the process $\tau \to \mu f_0(980)$ we substitute 
the quark bilinears of (\ref{quarkbilinears}) in
(\ref{amplitude_quarks}) and use 
(\ref{rotation}) and (\ref{scalarcurrents}). Notice that it is just the scalar
part in $\left[ \overline{q} P_{L,R} q \right]$, and not the pseudoscalar,
the one that contributes in the present case. 
We obtain:
\begin{equation}
 T_H(\tau \to \mu f_0(980)) \, = \, \sum_{p=h^0, H^0} \, c_p \, \overline{\mu} \, \tau \, , 
\end{equation}
where
\begin{eqnarray}
 c_p & = & \frac{g}{2 m_W} \frac{1}{2 m_{H_p}^2} \, 
\left( J_L^{(p)} + J_R^{(p)} \right) \left(H_R^{(p)} + H_L^{(p)} \right) \, ,
\end{eqnarray} 
and
\begin{eqnarray}
 J_L^{(p)} & = &
\frac{c_m}{\sqrt{3}}\, 
\left\{ \frac{\sigma_2^{(p)*}}{\sin \beta} \left[ \frac{1}{\sqrt{2}} \sin \theta_S + 
 \cos \theta_S \right] \, m_{\pi}^2  \right. \, \nonumber \\
&& \; \; \; \; \; \; \;\; \;  \;- \, \left. \frac{\sigma_1^{(p)*}}{\cos \beta} \, \left[
\frac{3}{\sqrt{2}} \sin \theta_S \, m_{\pi}^2 + 
\left(   \cos \theta_S - 
\sqrt{2} \sin \theta_S \right) \, 2 \, m_K^2 \, \right] \right\} \, , \nonumber \\
J_R^{(p)} & = & J_L^{(p)*} \, .
\label{Jf0}
\end{eqnarray}
Notice that due to the
mass relations in (\ref{mq-mP}), the couplings
of the Higgs bosons, $h^0$ and $H^0$, to the quarks ($q=u,d,s$ ),
$S_{L,q}^{(p)}$ and $S_{L,q}^{(p)}$ in (\ref{Hqqcouplings}),  being proportional to 
the
quark masses, lead to Higgs-$f_0$ couplings that are proportional
to $m_P^2$ ($P=\pi, K$). This is seen clearly in the predicted functions 
$J_{L,R}^{(p)}$ of ~(\ref{Jf0}). In consequence, the dominant contributions to 
BR($\tau \to \mu f_0(980)$) will come clearly from the terms in the amplitude that are proportional to 
$m_K^2$.     

Finally, the result of the branching ratio for the $\tau \to \mu f_0(980)$ decay is
given by,

 \begin{equation} \label{eq:fullresult1}
 {\rm BR}(\tau \to \mu f_0(980))\, = \, \frac{1}{4 \pi} \,
 \frac{\lambda^{1/2}(m_{\tau}^2,m_{\mu}^2,m_{f_0}^2)}{m_{\tau}^2 \Gamma_\tau} \,
\frac{1}{2} \, \sum_{i,f} |T_H|^2 \,, 
\end{equation}
where
\begin{equation}\label{eq:fullresult2}
 \frac{1}{2} \sum_{i,f} |T_H|^2 \, = \,  \frac{\left(m_{\mu}+m_{\tau}\right)^2 -
 m_{f_0}^2}{4 \, m_{\tau}} \, |c_{h^0} + c_{H^0} |^2 \, , 
\end{equation}
$\Gamma_\tau$ is the total $\tau$ width and 
$\lambda(x,y,z) = (x+y-z)^2-4xy$.

\noindent 
{\bf Approximate formula}

Next we derive a simple formula which approximates 
reasonably well our full one-loop prediction in (\ref{eq:fullresult1}) and
(\ref{eq:fullresult2}). For this, we work within the 
approximation of large $\tan \beta$ that is appropriate for LFV tau decays,
whose rates grow quite fast with this parameter. This is especially relevant
for channels where the LFV rates are dominated by the Higgs mediated diagrams,
as it is the present case, and where the growth with $\tan \beta$ is extremely 
pronounced. 

The other approximation which is used frequently in the
literature, due to its simplicity, is the use of the mass insertion (MI) 
method, where 
the tau-muon
LFV is encoded in the dimesionless parameters $\delta_{32}^{XY}$
($XY=LL,RR,LR$). In the SUSY models
the dominant one is $\delta_{32}^{LL}$ and its expression in the LLog
approximation, $(\delta_{32}^{LL})_{\rm LLog} \equiv \delta_{32}$, is that 
given in (\ref{delta32_Llog}).

It is known~\cite{Arganda:2004bz}~\cite{Arganda:2005ji} that 
at large $\tan\beta$ the vertex function $H_L$ dominates $H_R$ by about a factor 
$m_\tau /m_\mu$. In addition
 $H_L^{H^0}$ is by far larger than $H_L^{h^0}$
in this limit, and one can safely neglect the later one. 
More specifically, by using the MI approximation, 
its chargino and neutralino contributions in the large $\tan\beta$ and
heavy $M_{\rm SUSY}$ limits
give, correspondingly, the following expressions~:
\begin{eqnarray}
  H_{L,c}^{(H^0)}\,&=&\,  \,\frac{g^3}{16\pi^2} \,
\frac{m_\tau}{12 m_W} \, \delta_{32} \, \tan^2{\beta} \, ,\nonumber \\[2mm]
  H_{L,n}^{(H^0)}\,&=&\,  \,\frac{g^3}{16\pi^2} \,
\frac{m_\tau}{24 m_W} \, (1-3 \tan^2{\theta_W}) \,\delta_{32} \, \tan^2{\beta}  \, .
\label{HL_semilept}
\end{eqnarray}
One can further verify that $H_c$ dominates $H_n$ by about a factor 20, so
that we will simplify $H_L \simeq H_{L,c}$.

On the other hand, we also consider the
large $\tan\beta$ limit of the functions that define the 
$H^0$ couplings to $f_0(980)$, $J_L$ and $J_R$ in (\ref{Jf0}).
We obtain~:

\begin{eqnarray}\label{JH0approx}
J_L^{(H^0)} &=& J_R^{(H^0)} \, = \, \frac{F}{2\sqrt{3}} \tan\beta 
\left[ \frac{3}{\sqrt{2}}  \sin\theta_S \,  m_\pi^2+ 
( \cos\theta_S- \sqrt{2}\sin\theta_S) 2 m_K^2 \right]. 
\end{eqnarray}

By using the above sequence of approximations and by neglecting the
muon mass, we finally get the following
simple result:
\begin{eqnarray}\label{taumuf0_approx}
\text{BR}(\tau \to \mu f_0(980))_{\text{approx}}  &=& 
\frac{1}{16 \pi m_\tau^3} \left( m_\tau^2 -
m_{f_0}^2 \right)^2 \left| \frac{g}{2 m_W} \, \frac{1}{m_{H^0}^2} \, 
J_L^{(H^0)} \, H_{L,c}^{(H^0)} \right|^2
\frac{1}{\Gamma_\tau }  \\[2mm]
&=& \! \! \! \left( \!  \begin{array}{c} 7.3 \times 10^{-8}\,\, (\theta_S=7^\circ) \\ 
4.2 \times 10^{-9}\,\, (\theta_S=30^\circ)  \end{array} \! \! \right) 
\left| \delta_{32} \right|^2    \left( \frac{100}
{m_{H^0}({\rm GeV})}
\right)^4 \left( \frac{\tan \beta}{60} \right)^6  \! \! . \nonumber
\end{eqnarray}
In the last line we see explicitly the fast growth with $\tan \beta$, as
$(\tan \beta)^6$, the expected dependence with the relevant Higgs mass, as 
$(m_{H^0})^{-4}$, and also with the LFV parameter, as $|\delta_{32}|^2$. The 
two numerical factors correspond to the two assumed values 
for the mixing angle that defines the $f_0(980)$ state,  
$\theta_S= 7^\circ$ and $\theta_S= 30^\circ$. These two results differ by a  
factor 17, meaning that the predicted rates will carry a theoretical uncertainty
of about this number, due to the uncertainty in the definition of the $f_0(980)$
state.   

\noindent
{\bf Numerical results}

In the following we present the numerical predictions for 
BR($\tau \to \mu f_0(980)$). 
We first show the results from the full computation in (\ref{eq:fullresult1}) and 
(\ref{eq:fullresult2}) and then compare with the
approximate results in (\ref{taumuf0_approx}) and also with the rates of other 
LFV tau decay channels.

\begin{figure}[h!]
   \begin{center} 
        \begin{tabular}{cc} \hspace*{-12mm}
  \psfig{file=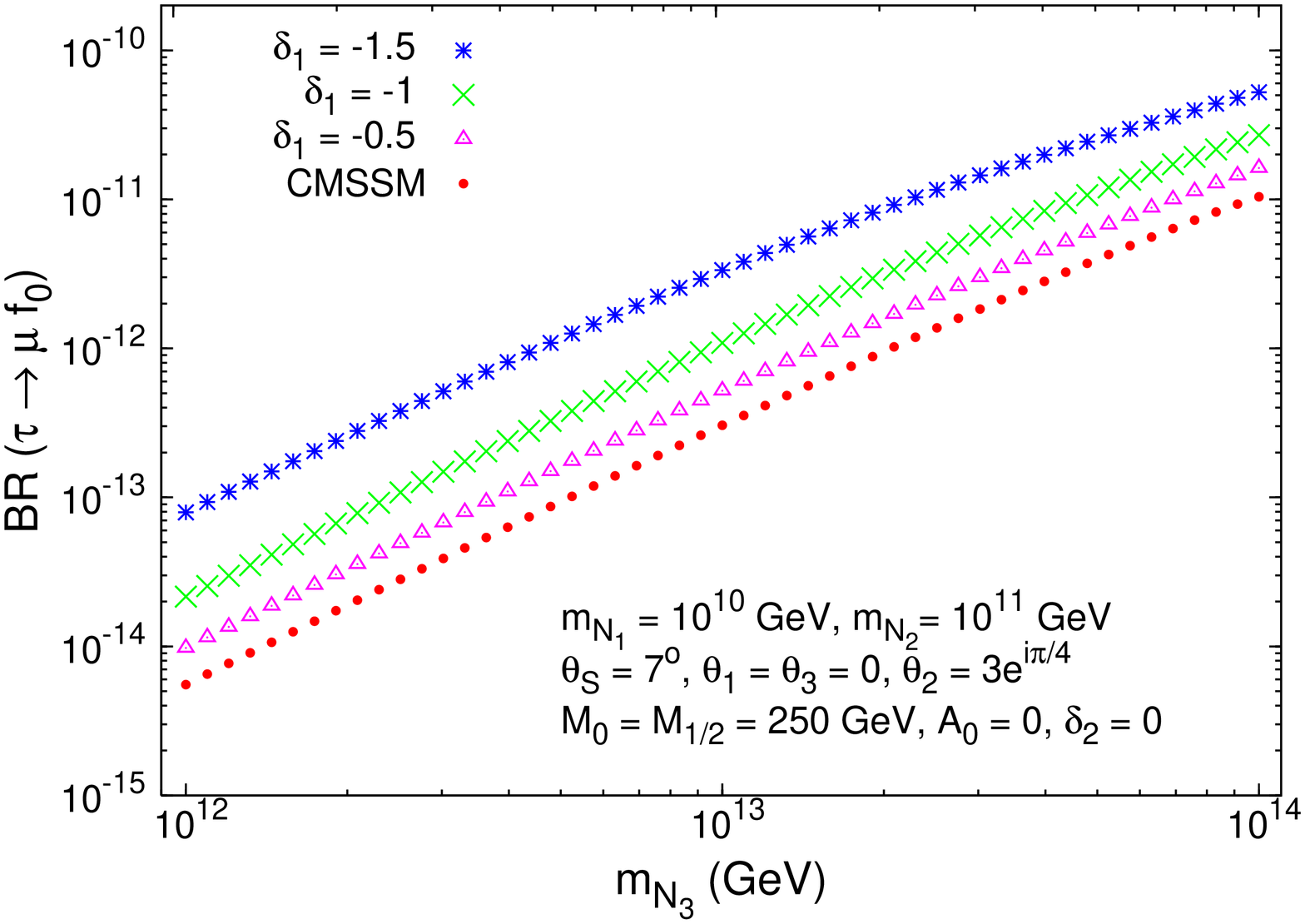,angle=270,width=85mm,clip=} 
&
  	\psfig{file=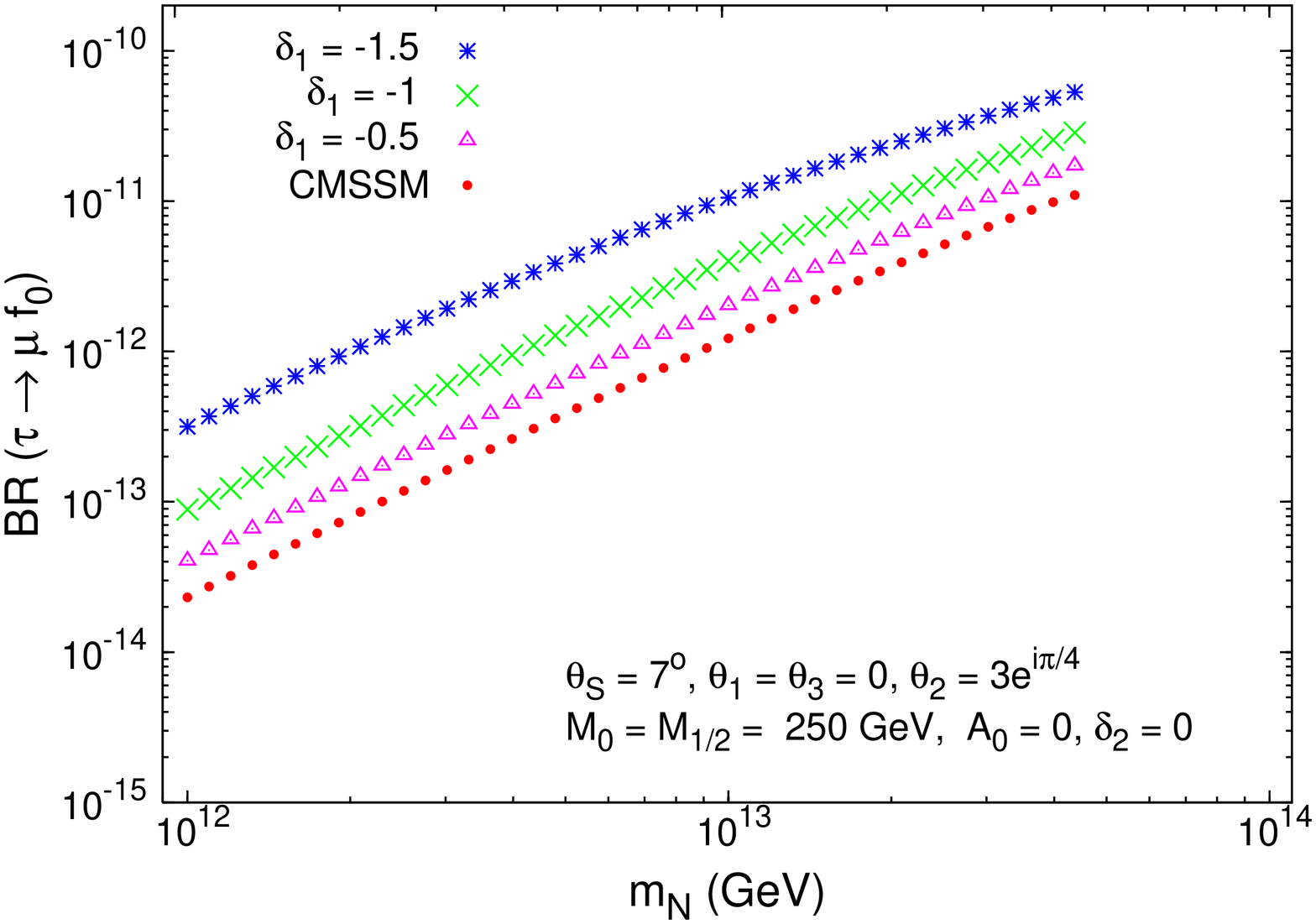,angle=270,width=85mm,clip=}  
     \end{tabular}
     \caption{BR($\tau \to \mu f_0(980)$) in the NUHM-Seesaw, for several $\delta_1$
     values, and in the CMSSM-Seesaw 
     versus the relevant heavy neutrino mass, 1) for hierarchical heavy 
     neutrinos (left panel), and 2) degenerate heavy neutrinos (right panel).
  } \label{fig:BRf0_mNi} 
   \end{center}
 \end{figure}

In Fig.~\ref{fig:BRf0_mNi} it is shown the  
BR($\tau \to \mu f_0(980)$) versus the heavy neutrino masses, in both
scenarios with hierarchical and degenerate heavy neutrinos. In the hierarchical
case we display just the dependence with the relevant mass, $m_{N_3}$.
 As expected, from the previously manifested
behaviour of 
$|\delta_{32}|$ with $m_{N_3}$ (or with $m_N$, in the degenerate case) in 
Fig.~\ref{fig:contours}, we find a fast growing of  BR($\tau \to \mu f_0(980)$)
with this mass.
Although not explicitely shown here, we have also checked
in the hierarchical case, the near independence on
the other masses, $m_{N_1}$ and $m_{N_2}$.  
From this figure it is also evident that by choosing properly the 
$\delta_1$ and $\delta_2$ parameters
of the NUHM scenario, such that the relevant Higgs boson mass $m_{H^0}$ 
gets lower than for $\delta_1=\delta_2=0$, 
the branching ratios get larger than 
in the CMSSM scenario. Finally, 
by comparing the rates of the two neutrino scenarios, and for the same input
model parameter values, including the same $m_N$ and $m_{N_3}$,  we find  
rates in the degenerate case that are generally larger than in the 
hierarchical case. For instance, for the choice of input parameters in 
Fig.~\ref{fig:BRf0_mNi} we find larger rates by a factor of about 3. 
 In the following we will focus more on the hierarchical
case since it has the appealing feature of providing successful
baryogenesis, via leptogenesis, for some regions of the heavy neutrinos 
parameter space.   
 
\begin{figure}[h!]
   \begin{center} 
     \begin{tabular}{cc} \hspace*{-12mm}
  	\psfig{file=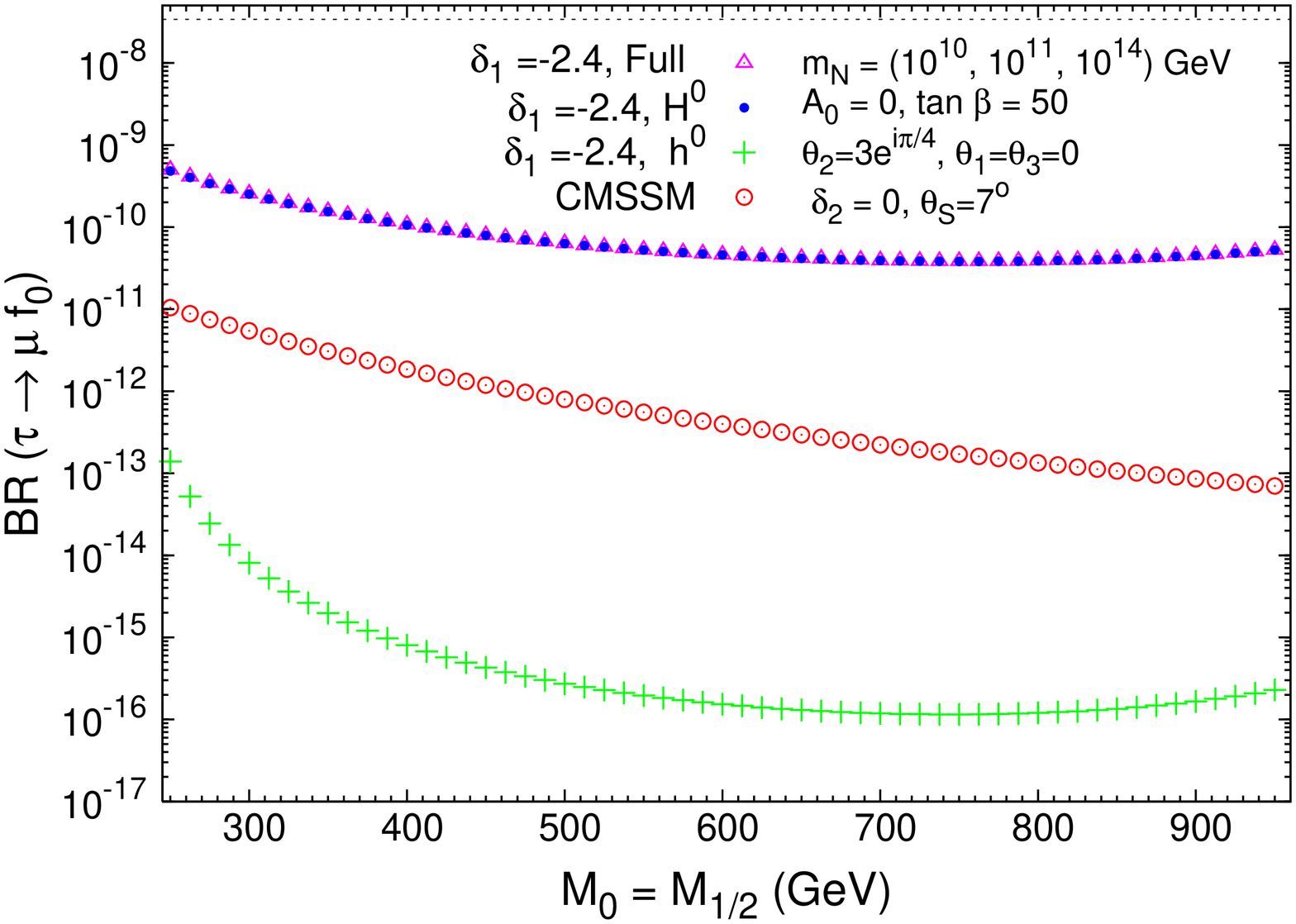,width=60mm,angle=270,clip=} 
	&
        \psfig{file=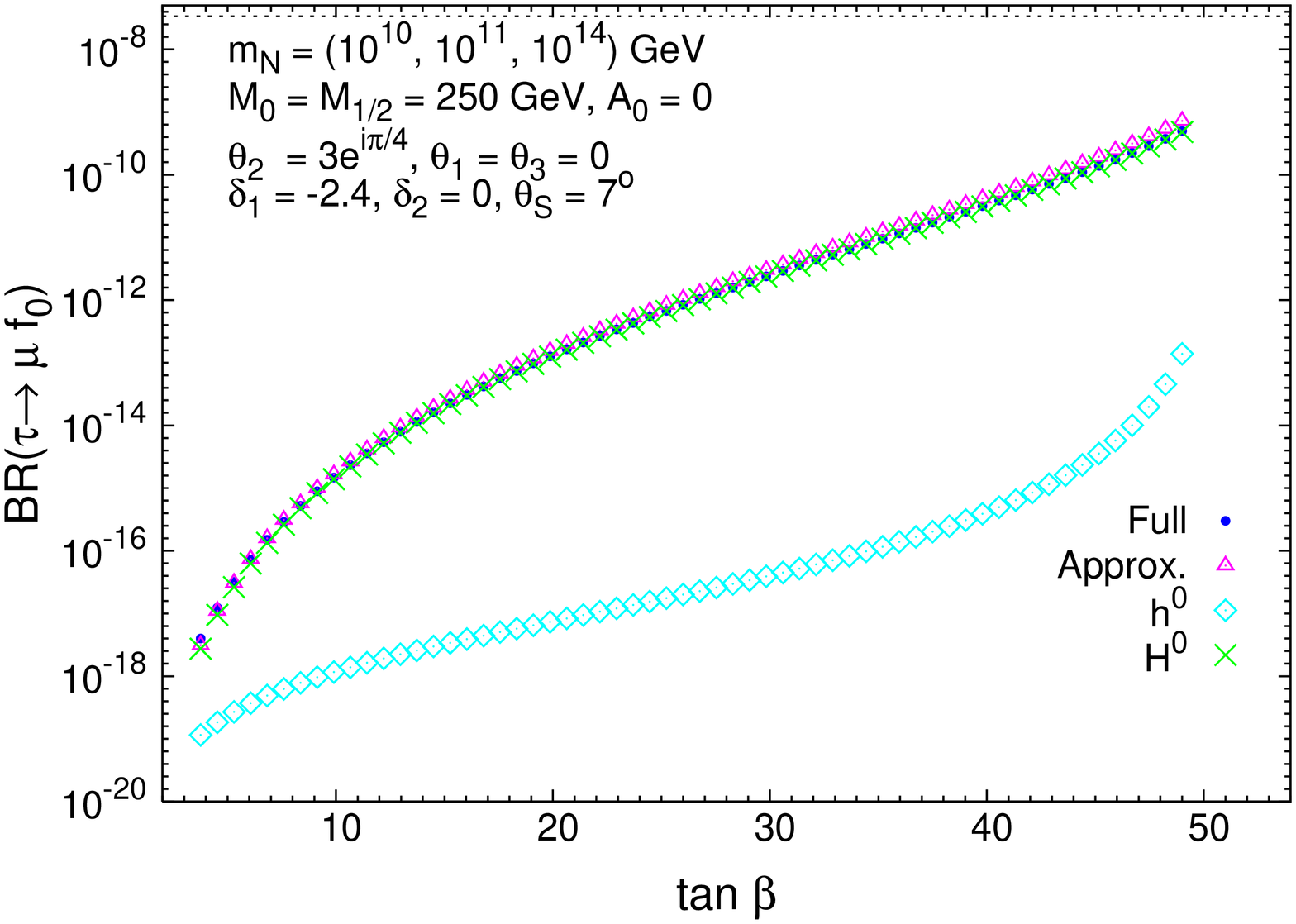,width=60mm,angle=270,clip=} 	  
       \end{tabular}
     \caption{BR($\tau \to \mu f_0(980)$) in the NUHM-Seesaw
     scenario: 1) As a function of   $M_0=M_{1/2}=M_{\rm SUSY}$ (left panel). We
     show separately  the $H_0$ and $h_0$ contributions as well as the total.
  The predictions for the total rates within the CMSSM-Seesaw scenario are also included 
  for comparison;
  2) As a function of  $\tan\beta$ (right panel). Again, the dominant $H^0$,
  the subdominant $h^0$ and the 
  total rates are displayed. We also include here the approximate predictions given by 
  (\ref{taumuf0_approx}) for comparison with the full rates. The dotted horizontal line
  at the top of the graphics is the present experimental upper bound.
 } \label{fig:BRf0_msusy_tanbeta} 
   \end{center}
 \end{figure}

 We present the predictions of the BR($\tau \to \mu f_0(980)$) versus de soft
SUSY masses 
$M_0$ and $M_{1/2}$ in Fig.~\ref{fig:BRf0_msusy_tanbeta}.  Here we take again
 $M_0=M_{1/2} \equiv M_{\rm SUSY}$ and compare the results 
 in both scenarios, the NUHM with $\delta_1=-2.4$ and $\delta_2=0$, where the
 predicted Higgs boson masses for large $\tan \beta \sim 50$ lay 
 within the interval 100-250 GeV, and the
 CMSSM. The most evident feature in this plot is the different behaviour of the 
 BR($\tau \to \mu f_0(980)$) with $M_{\rm SUSY}$ in these two scenarios. Whereas in
 the CMSSM the rates are found to decrease with increasing  
 $M_{\rm SUSY}$, as expected, it clearly does not happen in the NUHM. In fact, the rates are
 practically constant for $M_{\rm SUSY}> 400$ GeV.  The reason for this behaviour
 is that the SUSY 
particles do not decouple at large $M_{\rm SUSY}$ in this decay. 
The non-decoupling behaviour 
can be checked
analytically in that the LFV vertex, described by the dominant 
form factor
$H_L$,
 tends to a constant value at asymptotically 
large $M_{\rm SUSY}$, as indicated in (\ref{HL_semilept}).
Since, on the other hand,  $m_{H^0}$ is kept at the low region 
even for  
large $M_{\rm SUSY}$, then a constant $H_L$ with $M_{\rm SUSY}$  implies
 approximately constant BR($\tau \to \mu f_0(980)$) as well. 
 
Another interesting feature of the predicted rates in the NUHM scenario, 
that is manifested in Fig.~\ref{fig:BRf0_msusy_tanbeta} as well,    
 is the clear dominance by many orders of magnitude
of the $H^0$ contribution over the $h^0$ one in the whole $M_{\rm SUSY}$ 
considered interval. 
This is due to the fact that at large $\tan \beta$ the $H^0$ contribution 
is enhanced by a $\tan^6 \beta$ factor, whereas the $h^0$ one is suppressed in
this limit. In fact, we also see in this plot that the total rates are nearly 
indistinguishable from the $H^0$ contributions. Thus, to neglect the $h^0$
contribution is an extremely good approximation. 

 
Concerning the Higgs sector parameters, the BR($\tau \to \mu f_0(980)$) 
is mainly sensitive to $\tan \beta$ and $m_{H^0}$ since, as said before,
the $H^0$-mediated LFV semileptonic decays grow very fast with both 
$\tan \beta$ and  $1/m_{H^0}$. In fact, 
in the approximation given in 
(\ref{taumuf0_approx}), as already said, BR($\tau \to \mu f_0(980)$)  
goes as  $(\tan\beta)^6$ and $(1/m_{H^0})^4$, respectively.

The predictions of BR($\tau \to \mu f_0(980)$) as a function of  
$\tan \beta$ are shown in the right panel of Fig.\ref{fig:BRf0_msusy_tanbeta}. 
We show again separately the
$h^0$ and $H^0$ contributions and the total rates which are clearly dominated
by the $H^0$ in the full studied interval of $\tan \beta$. Besides, it also 
displays 
the fast
growing of the total rates with $\tan \beta$ , reaching values at the
$\sim 10^{-9}$ level for  $\tan \beta \sim 50$ which are close but still 
below the present experimental bound.
We also see that the particular shape of the curve for the total rates 
is a consequence as
well of the $m_{H^0}$ dependence with $\tan \beta$ in these SUSY scenarios,
as illustrated in Fig.~\ref{fig:mH_CMSSM-NUHM}.

The comparison between our predictions for the full result in 
(\ref{eq:fullresult1}) and (\ref{eq:fullresult2}) and the 
approximate result in (\ref{taumuf0_approx}), 
which includes just the $H^0$ boson contribution, can be seen as well
in Fig.~\ref{fig:BRf0_msusy_tanbeta}.  
The agreement between the full and the approximate 
results is quite remarkable, for all the studied values in the
$5 \leq \tan \beta\leq 50$ range. Therefore, we conclude that our simple formula
(\ref{taumuf0_approx}) provides a very good approximation to 
BR($\tau \to \mu f_0(980)$) for all $\tan \beta$.

It is interesting to compare $\tau \to \mu f_0(980)$ to other Higgs-mediated LFV 
tau decay channels  
like $\tau \to \mu \eta$ and $\tau \to 3 \mu$. First, notice that
our previous result of the $H^0$ dominance 
in  the $\tau \to \mu f_0(980)$ channel over the full $\tan \beta$ interval, 
is not true for  
the correlated channel 
 $\tau \to \mu \eta$, nor the leptonic $\tau \to 3 \mu$ decay.  
The semileptonic LFV $\tau \to \mu \eta$ decay  can be
mediated by a CP-odd $A^0$ Higgs boson and a Z boson, but the 
contribution from $A^0$ dominates the full rates only in the large  
$\tan \beta \geq 20$ region~\cite{Arganda:2008jj,Arganda:2008yc}.  The
$\tau \to 3 \mu$ channel 
can be mediated (apart from the box diagrams, which are negligible) by a photon, a Z boson and the three neutral 
Higgs bosons, $h^0$, $H^0$ and $A^0$~\cite{Arganda:2005ji}. The photon dominates 
largely this decay,
except at the extreme high values of $\tan\beta \geq 60$ and $M_{SUSY} \geq 1$ 
TeV, where the two type of contributions from the photon and the Higgs bosons, 
$H^0$ and $A^0$ compete.
These features can be seen clearly by comparing the corresponding 
approximate formulas, valid at large $\tan \beta$, for their respective 
Higgs boson 
contributions. 
That is, one should compare our result in (\ref{taumuf0_approx}) 
to the previous results of BR($\tau \to \mu
\eta$)~\cite{Brignole:2004ah,Arganda:2008jj} and 
BR($\tau \to 3
\mu$)~\cite{Babu:2002et,Dedes:2002rh,Brignole:2003iv,Arganda:2005ji} 
for the same input parameters. These are~\cite{Arganda:2008jj},
\begin{eqnarray}
\text{BR}(\tau \to \mu \eta)_{H_\text{approx}}  &=& \frac{1}{8 \pi m_\tau^3} \left( m_\tau^2 -
m_\eta^2 \right)^2 \left| \frac{g}{2 m_W} \, \frac{F}{m_{A^0}^2} \, 
B_L^{(A^0)}(\eta) \, H_{L,c}^{(A^0)} \right|^2
\frac{1}{\Gamma_\tau } \nonumber \\[2mm]
&=& 1.2 \times 10^{-7} (\theta= -18^o) \left| \delta_{32} \right|^2 \left( \frac{100}
{m_{A^0}({\rm GeV})}
\right)^4 \left( \frac{\tan \beta}{60} \right)^6,
\label{taumueta_approx}
\end{eqnarray}
where,
\begin{equation}
B_L^{(A^0)}(\eta) \, = \, -i\frac{1}{4\sqrt{3}} \tan\beta 
\left[ (3 m_\pi^2 - 4 m_K^2) \cos\theta - 
2 \sqrt{2} m_K^2 \sin\theta \right] \, , \, H_{L,c}^{(A^0)}= i H_{L,c}^{(H^0)}
\,,
\end{equation} 
and: 
\begin{eqnarray} \!\!\!\!\!\!\!\!\!\!
\text{BR}(\tau \to 3 \mu)_{H_\text{approx}}  &=& 
\frac{G_F^2}{2048 \pi^3} \frac{m_\tau^7 m_\mu^2}{\Gamma_\tau} 
\left( \frac{1}{m_{H^0}^4} + \frac{1}{m_{A^0}^4} + \frac{2}{3m_{H^0}^2 m_{A^0}^2} \right) 
\left| \frac{g^2\delta_{32} }{96 \pi^2} \right|^2 (\tan\beta)^6 
\label{tau3mu_approx_formula}\\
 &=& 1.2 \times 10^{-7} \left| \delta_{32} \right|^2 \left( \frac{100}{m_{A^0}({\rm GeV})}
 \right)^4 \left( \frac{\tan \beta}{60} \right)^6.
\label{tau3mu_
H_approx} 
\end{eqnarray}
From this comparison, we conclude that, for the same choice of the
model parameters, and for
$\theta_S=7^\circ$, the three rates 
BR($\tau \to \mu f_0(980)$), BR($\tau \to \mu \eta$) and BR($\tau \to 3 \mu$) 
are very similar if $\tan \beta \gtrsim 60$ and $M_{SUSY} \gtrsim 1$ TeV.
Concretely, we predict 
BR($\tau \to \mu f_0(980)$):BR($\tau \to 3 \mu$):BR($\tau \to \mu \eta$) 
$\sim 0.6:1:1$, and   
they are all at the 
$\sim {\cal O}(10^{-7})$ level for $|\delta_{32}| \sim 1$, 
$m_H \sim 100$ GeV and $\tan \beta \sim 60$. 
Therefore, the three are closely competitive channels. It should also be
mentioned that our estimate
of BR($\tau \to \mu f_0(980)$) for $\theta_S \simeq 7^o$ and for the same 
input parameters, 
$m_H$, $\tan \beta$ and $|\delta_{32}|$, is about one order of
magnitud smaller than the prediction in \cite{Chen:2006hp}. They also predict
a different ratio among the three LFV channels of $\sim 1.3:0.5:1$. We believe
that the main differences come from our different approaches for hadronization
which produce, as we have already said, a dispersion in the results by 
a factor of ${\cal O}(10)$.

\begin{figure}[h!]
   \begin{center} 
     \begin{tabular}{c} \hspace*{-12mm}
  	\psfig{file=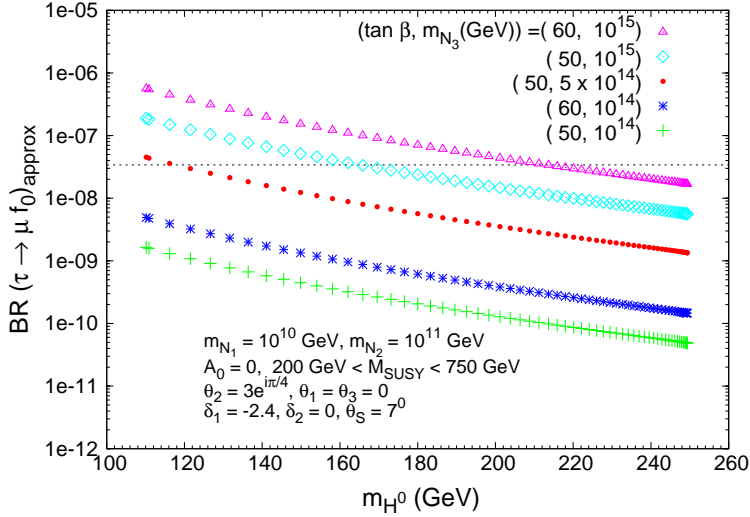,width=70mm,angle=270,clip=}   
       \end{tabular}
     \caption{Sensitivity to the Higgs Sector in $\tau \to \mu f_0(980)$ within
      the NUHM-Seesaw
     scenario. The predicted rates are within the approximation of 
     (\ref{taumuf0_approx}) and are displayed  
     as a function of 
     $m_{H^0}$,  for various choices of large $m_{N_3}$ and $\tan \beta$. The dotted horizontal line
  is the present experimental upper bound.
  } \label{fig:finalplot} 
   \end{center}
 \end{figure}

\begin{figure}[h!]
   \begin{center} 
     \begin{tabular}{c} \hspace*{-12mm}
  	\psfig{file=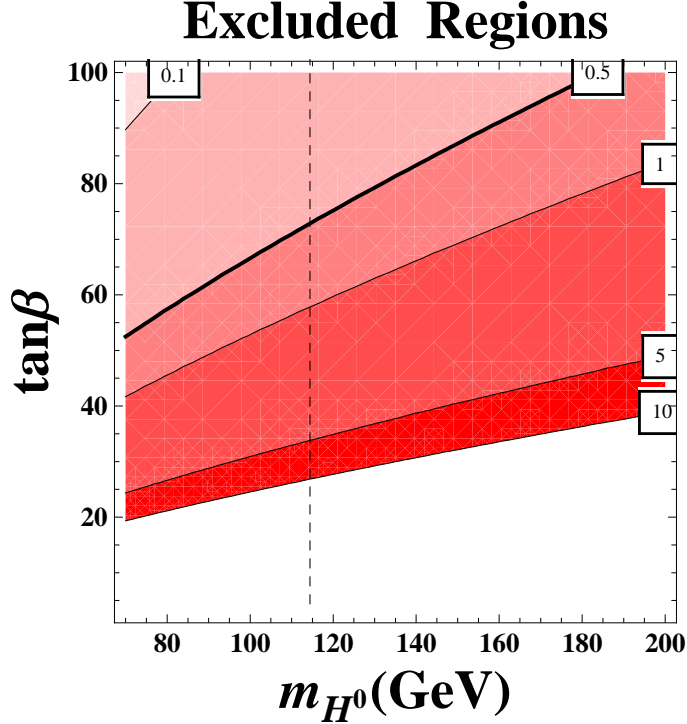,width=90mm,clip=}   
       \end{tabular}
     \caption{The excluded regions in the ($m_{H^0}$, $\tan \beta$) plane are 
       the areas above the 
     contour lines corresponding to fixed $|\delta_{32}|=0.1,0.5,1,5,10$.}  
   \label{exclusion_regions} 
   \end{center}
 \end{figure}

Finally, we summarize the sensitivity to the Higgs sector in the NUHM-Seesaw
scenario in Fig.\ref{fig:finalplot}. In this plot we are using the approximate
formula in (\ref{taumuf0_approx}) and we are 
setting $\theta_2=3\, e^{i \frac{\pi}{4}}$ and $\delta_1=-2.4$, 
$\delta_2=0$. The soft masses are varied in the range
$200 \,{\rm GeV}\,\leq M_0=M_{1/2} \equiv M_{\rm SUSY}\,\leq 750\,{\rm GeV}$. The explored $m_{H^0}$
values in this plot correspond precisely to the output Higgs masses for this
later $M_{\rm SUSY}$ interval. The main conclusion from this plot is that for 
large $m_{N_3} \sim 5 \times 10^{14}-10^{15}$ GeV and large 
$\tan \beta \sim 50-60$ the predicted rates are already at the present experimental
reach and, therefore, there is indeed Higgs sensitivity in this channel. 
In this concern, we find interesting to further explore if with the
present experimental bound of 
BR$(\tau \to \mu f_0(980)) \times $ BR$(f_0(980) \to \pi^+ \pi^-)$ $<\, 3.4\times10^{-8}$ 
one may already 
exclude some region of the model parameter space. Our conclusion is that indeed
it is possible to exclude the regions in the ($m_{H^0}$, $\tan \beta$) plane 
as summarized in Fig.~\ref{exclusion_regions}. 
In this plot we assume, for simplicity, 
BR$(f_0(980) \to \pi^+ \pi^-) \sim 1$ and choose the specific input values,
$|\delta_{32}|=0.1,0.5,1,5,10$. For each fixed  $|\delta_{32}|$ the excluded
region is the area above the corresponding contour line.  
For completeness,  we have also included in this plot the present experimental lower 
bound for the SM Higgs mass at 114.4 GeV. Some words of caution should 
be said, anyway, about
the conclusions from this plot since there are large uncertainties involved 
in the theoretical estimate of BR$(\tau \to \mu f_0(980))$. There are two main
ones: 1) the uncertainty in the definition of $f_0(980)$ that, as evaluated 
in (\ref{taumuf0_approx}), can produce 
a dispersion of more than one order of magnitude in the predicted rates, 
and 2) the use of the approximate formula for values of $|\delta_{32}|>0.5$ which
are out of the region that is allowed by a perturbative approach. The use of the
MI approximation for such large values of $|\delta_{32}|$ is also questionable.   

\section{Conclusions} 
In this work we have studied in full detail the LFV semileptonic tau decay 
channel $\tau \to \mu f_0(980)$ within the context of two constrained SUSY-Seesaw
models, the CMSSM-Seesaw and the NUHM-Seesaw which have very different Higgs
sector spectra. Concretely, we have selected NUHM-Seesaw scenarios 
with a light Higgs sector, $h^0$, $H^0$ and $A^0$, in the 100-250 GeV
range, and considered several possibilities for the SUSY sector, varying
the SUSY mass $M_{\rm SUSY}$ in the 200-1000 GeV range.
Through all this analysis, 
 we have required compatibility  
with both the present experimental upper bound for this decay and 
with neutrino data for masses and oscillations.

We have presented a full computation of BR$(\tau \to \mu f_0(980))$ that 
includes 
the complete one-loop SUSY diagrams with charginos, neutralinos, sleptons 
and
sneutrinos contributing in the loops to the relevant effective LFV 
$\tau \mu H$ vertex. We have also taken into account the two kind of 
Higgs-mediated 
diagrams, with $h^0$ and $H^0$ in the internal propagator connecting the 
LFV
vertex with the final quark-antiquark pairs. On the other hand, and
in order to provide predictions for the final meson $f_0(980)$, we have 
performed the hadronization of
the quark bilinears by means of the standard techniques in $\chi$PT and
R$\chi$T. We have shown that in this chiral approach, 
the Higgs coupling to the $f_0(980)$ is dominated by its strange quark 
component. The leading term in this coupling is proportional to $m_K^2$, 
which is a consequence of the Gell-Mann--Oakes--Renner mass relation
($B_0m_s=m_K^2-1/2m_\pi^2$), and the fact that the Higgs
coupling to the strange quark is proportional to $m_s$. On the other hand, the
$H^0-f_0$ coupling is dominant over the $h^0-f_0$ coupling since the first one
goes approximately as $\tan \beta$ in the large $\tan \beta$ limit 
(due again to this behaviour of the $Hss$ 
coupling), whereas the
second one is suppressed in this limit.

We have analysed in full detail the dependence of
BR$(\tau \to \mu f_0(980))$ with all the parameters defining the two  
constrained SUSY-Seesaw scenarios and we have extracted from this analysis
which are the relevant ones. Regarding the heavy neutrino sector, and for the
most BAU favorable scenario of hierarchical heavy neutrinos, the most relevant
parameters are the heaviest neutrino mass, $m_{N_3}$ and the $\theta_{1,2}$
angles. Concerning the SUSY and Higgs sectors the most relevant 
parameters are the SUSY masses, driven by $M_{\rm SUSY}$, the CP-even 
Higgs boson mass, $m_{H^0}$, and $\tan \beta$. 

In the numerical predictions, we have found much larger
rates in the NUHM-Seesaw than in the CMSSM-Seesaw scenarios, due mainly to the 
lighter Higgs mass $m_{H^0}$ found in the first scheme. Indeed, 
it is just in the NUHM-Seesaw case where the predictions for 
BR$(\tau \to \mu f_0(980))$ can
reach the present experimental sensitivity. We have shown, that in order to
get values of BR$(\tau \to \mu f_0(980))$ at the $10^{-8}-10^{-7}$ level one needs 
large values for the relevant parameters, namely, $m_{N_3} \sim 10^{14}-10^{15}$ GeV, 
$|\theta_{1,2}| \sim 2-3$, $\pm$ arg$(\theta_{1,2})\sim \pi/4-3\pi/4$,
$\tan \beta \sim 50-60$ and $m_{H^0} \sim 100-200$ GeV.  

In addition to the full results, we have provided an approximate simple formula
for BR$(\tau \to \mu f_0(980))$ which has been obtained in the large $M_{\rm SUSY}$
and large $\tan \beta$ limit, and with the MI approximation for the relevant LFV parameter 
$\delta_{32}$. Furthermore, we have shown in this work that this approximate 
result
agrees pretty well with the full result in practically all the explored
parameter space. The main basic features of the full predicted rates 
are very well reproduced by the simple formula in (\ref{taumuf0_approx}),
which summarizes the fast growing with $\tan \beta$, going as $(\tan\beta)^6$, 
with $1/m_{H^0}$, going as $(1/m_{H^0})^4$, and being approximately constant 
with $M_{\rm SUSY}$. The dependences with $m_{N_3}$ and $\theta_{1,2}$ go
via the $\delta_{32}$ parameter, and the large $m_{N_3}$ values are
what enhance dominantly the
rates, growing approximately as BR $\sim |m_{N_3} \log m_{N_3}|^2$. 

The most important conclusion from this work, as illustrated in 
Figs.~\ref{fig:finalplot} and \ref{exclusion_regions}, is that the 
LFV tau decay $\tau \to \mu f_0(980)$ is indeed
sensitive to the Higgs sector of the NUHM-Seesaw models. Concretely, it
is mostly sensitive to the CP-even Higgs boson $H^0$, and therefore it
complements very nicely the previous searches via the $\tau \to \mu \eta$ decay
 which is sensitive to the CP-odd Higgs boson $A^0$. These two channels together
 with the leptonic $\tau \to 3 \mu$ decay are
 undoubtly the most competitive LFV tau decays where to look for indirect 
 Higgs signals. As a final product of our analysis we have extracted some excluded 
 areas in the
 parameter space of these models by using our approximate formula.  
 The sensitivity found here to the Higgs sector  
 will presumably improve in the future if the experimental
 reach increases up to $10^{-9}-10^{-10}$, as it seems to be the case in the future
 SuperB and flavour factories \cite{Browder:2008em}.  

\noindent
{\bf Acknowledgments}

M.J.Herrero wishes to thank Simon Eidelman for his encouragement to make this
work and for interesting discussions on LFV. 
This work has been supported in part by the EU
   MRTN-CT-2006-035482 (FLAVIAnet), by MEC (Spain) under grants
   FPA2006-05423 and FPA2007-60323, by Generalitat Valenciana under the excellence grant
   PROMETEO/2008/069, by Comunidad de Madrid under HEPHACOS project and
   by the Spanish Consolider-Ingenio 2010 Programme CPAN (CSD2007-00042).
   A.M.Rodriguez-Sanchez acknowledges MEC for her FPU fellowship (AP2006-02535).



\end{document}